\documentclass[twocolumn]{aastex63}
\usepackage{graphicx}
\usepackage{float}

\newcommand{\xrt}{Swift/XRT}
\newcommand{\chandra}{Chandra}

\newcommand{\lsim }{{\lower0.8ex\hbox{$\buildrel <\over\sim$}}}
\newcommand{\gsim }{{\lower0.8ex\hbox{$\buildrel >\over\sim$}}}
\newcommand{\Msun}{\ifmmode {M_{\odot}}\else${M_{\odot}}$\fi}
\newcommand{\Lsun}{\ifmmode {L_{\odot}}\else${L_{\odot}}$\fi}
\newcommand{\Rsun}{\ifmmode {R_{\odot}}\else${R_{\odot}}$\fi}

\graphicspath{{./}{figures/}}

\shorttitle{Accreting neutron stars in globular clusters}
\shortauthors{Panurach et al.}

\begin{document}

\title{The MAVERIC Survey: Variable jet--accretion coupling in luminous accreting neutron stars in Galactic globular clusters}

\correspondingauthor{Teresa Panurach}
\email{panurach@msu.edu}

\author[0000-0001-8424-2848]{Teresa Panurach}
\affiliation{Center for Data Intensive and Time Domain Astronomy, Department of Physics and Astronomy, Michigan State University, East Lansing MI, USA}

\author[0000-0002-1468-9668]{Jay Strader}
\affiliation{Center for Data Intensive and Time Domain Astronomy, Department of Physics and Astronomy, Michigan State University, East Lansing MI, USA}

\author[0000-0003-2506-6041]{Arash Bahramian }
\affiliation{International Centre for Radio Astronomy Research Curtin University, GPO Box U1987, Perth, WA 6845, Australia}

\author[0000-0002-8400-3705]{Laura Chomiuk}
\affiliation{Center for Data Intensive and Time Domain Astronomy, Department of Physics and Astronomy, Michigan State University, East Lansing MI, USA}

\author[0000-0003-2506-6041]{James C. A. Miller-Jones}
\affiliation{International Centre for Radio Astronomy Research Curtin University, GPO Box U1987, Perth, WA 6845, Australia}

\author[0000-0003-3944-6109]{Craig O. Heinke}
\affiliation{Department of Physics, University of Alberta, CCIS 4-181, Edmonton, AB T6G 2E1, Canada}

\author[0000-0003-0976-4755]{Thomas J. Maccarone}
\affiliation{Department of Physics \& Astronomy, Texas Tech University, Box 41051, Lubbock, TX 79409-1051, USA}

\author[0000-0003-0286-7858]{Laura Shishkovsky}
\affiliation{Center for Data Intensive and Time Domain Astronomy, Department of Physics and Astronomy, Michigan State University, East Lansing MI, USA}

\author[0000-0001-6682-916X]{Gregory R. Sivakoff}
\affiliation{Department of Physics, University of Alberta, CCIS 4-181, Edmonton, AB T6G 2E1, Canada}

\author[0000-0002-4039-6703]{Evangelia Tremou}
\affiliation{LESIA, Observatoire de Paris, CNRS, PSL, SU/UPD, Meudon, France}

\author[0000-0003-4553-4607]{Vlad Tudor}
\affiliation{International Centre for Radio Astronomy Research Curtin University, GPO Box U1987, Perth, WA 6845, Australia}

\author[0000-0003-1814-8620]{Ryan Urquhart}
\affiliation{Center for Data Intensive and Time Domain Astronomy, Department of Physics and Astronomy, Michigan State University, East Lansing MI, USA}

\begin{abstract}

Accreting neutron stars in low-mass X-ray binaries show outflows---and sometimes jets---in the general manner of accreting black holes. However, the quantitative link between the accretion flow (traced by X-rays) and outflows and/or jets (traced by radio emission) is much less well-understood for neutron stars than for black holes, other than the general observation that neutron stars are fainter in the radio at a given X-ray luminosity. We use data from the deep MAVERIC radio continuum survey of Galactic globular clusters for a systematic radio and X-ray study of six luminous ($L_X > 10^{34}$ erg s$^{-1}$) persistent neutron star X-ray binaries in our survey, as well as two other transient systems also captured by our data. We find that these neutron star X-ray binaries show an even larger range in radio luminosity than previously observed. In particular, in quiescence at $L_X \sim 3 \times 10^{34}$ erg s$^{-1}$, the confirmed neutron star binary GRS 1747--312 in Terzan 6 sits near the upper envelope of the black hole radio/X-ray correlation, and the persistently accreting neutron star systems AC 211 (in M15) and X1850--087 (in NGC 6712) show unusual radio variability and luminous radio emission. We interpret AC 211 as an obscured ``Z source" that is accreting at close to the Eddington limit, while the properties of X1850--087 are difficult to explain, and motivate future coordinated radio and X-ray observations. Overall, our results show that neutron stars do not follow a single relation between inflow and outflow, and confirm that their accretion dynamics are more complex than for black holes. 
\end{abstract}

\section{Introduction} \label{sec:intro}
Neutron star low-mass X-ray binaries (LMXBs) contain a neutron star primary accreting through a disk from a $\lesssim 1 M_\odot$ secondary, which can be a low-mass main sequence star, an evolved star, or a degenerate star such as a white dwarf. Neutron star LMXBs are enormously overabundant (by a factor of $\sim 100$) in globular clusters, where they form dynamically due to the high stellar densities (e.g., \citealt{Clark75,Katz75,Ivanova08}).

The observed neutron star LMXBs in Galactic globular clusters include both persistent and transient sources \citep{Bahramian14,vandenBerg20}. The former are always X-ray luminous ($L_X \gtrsim 10^{35}$ erg s$^{-1}$), due to their high companion mass transfer rates, which keep the disk consistently in a hot, ionized state. The transient sources tend to have lower mass transfer rates from the secondary. This leads to the standard accretion disk ionization instability in which the disk alternates between a cool state with little accretion onto the neutron star (``quiescence") and a hot ``outbursting" state with substantial accretion \citep{Smak84,vanParadijs96,Lasota01}.

Since neutron stars have surfaces, it should be possible to trace the details of the accretion flow from the secondary to the disk, and thence as material accretes onto the neutron star, with some being ejected in an outflow or jet. X-ray observations primarily probe the inner regions of the accretion flow as well as the neutron star surface, while radio continuum data trace synchrotron emission associated with the outflow/jet.

Although extensive radio and X-ray observations have been made to constrain jet and accretion disk properties of all accreting compact objects (e.g., \citealt{Merloni03,Migliari06,Gallo18,Coppejans20,vandenEijnden21}), existing studies are biased towards black holes. Newly discovered transient X-ray sources in the field that can be accurately classified tend to be black holes rather than neutron stars, as black holes are on average brighter in the X-ray and far more radio-loud than neutron stars at nearly all accretion rates (e.g., \citealt{Corbel00,Migliari06,Tudor17,Gallo18}), with the caveat that some black holes and neutron stars deviate from this typical behavior (e.g., \citealt{Coriat11,Migliari11,Rushton16,Tudor17, Russell18, vandenEijnden18, Gusinskaia20b}). Despite the challenge, radio continuum studies of neutron stars are needed: there is a broad range of physical properties that are likely to affect the outflows/jets from these sources (such as their spin frequencies and surface magnetic field strengths), and there is strong observational evidence that accreting neutron stars show a greater variety of radio properties than accreting black holes (e.g., \citealt{Migliari11,Tudor17,Gusinskaia17,vandenEijnden21}).

One difficulty in studying field LMXBs is that their distances are generally unknown or poorly constrained \citep{Jonker04}, limiting the ability to draw accurate conclusions about the physics of the accretion. For example, distances inferred from X-ray bursts can have substantial systematic uncertainties (e.g., \citealt{Galloway08b}), and some ultracompact neutron star LMXBs do not show X-ray bursts at all. This issue is ameliorated in globular clusters, which have well-determined distances as well as a substantial population of neutron star LMXBs.

Here we present a radio and X-ray study of a sample of neutron star X-ray binary systems in Galactic globular clusters, focusing on eight luminous systems. In Section 2, we discuss the sample and observations. We use radio data from the Karl G. Jansky Very Large Array (VLA) or Australia Telescope Compact Array (ATCA) and we use X-ray data from  {Swift/X-Ray Telescope} (Swift/XRT) and the {Chandra X-ray Telescope} ({Chandra}). In Section 3, we consider the results from individual sources. We present discussion and a summary in Sections 4 and 5, respectively.

\section{Observational Data} 
\subsection{Our Sample}

Our sample includes all of the luminous ($L_X > 10^{34}$ erg s$^{-1}$) persistent LMXBs from the MAVERIC (Milky Way ATCA and VLA Exploration of Radio-sources in Clusters) radio continuum survey of 50 Galactic globular clusters (\citealt{Tremou18,Shishkovsky20}, Tudor et al.~2021, in prep). This survey includes all of the nearest ($\lesssim$ 9 kpc) massive ($\gtrsim 10^5$ M$_{\odot}$) clusters. 

The persistent LMXBs included in this study are: 4U 1746--37 (NGC 6441); 4U 1820--30 (NGC 6624); XB 1832--330 (NGC 6652); X1850--087 (NGC 6712); and AC 211 and M15 X-2 (M15). These represent 6 of the 8 persistent globular cluster sources known, excluding only 4U 0513--40 (NGC 1851; \citealt{Zurek09}) and 4U 1722--30 (Terzan 2; \citealt{intZand07}) as their host clusters are not in MAVERIC and have no deep ATCA or VLA data. To this initial sample we also add two other luminous transient sources. The first is GRS 1747--312 (Terzan 6), which shows approximately biannual outbursts \citep{Pavlinsky94,intZand03a}. The other is M15 X-3, an X-ray transient with limited existing studies of its radio properties \citep{Heinke09b,Strader12b} and one of two globular cluster LMXBs that has been observed at $L_X \sim 10^{34}$ erg s$^{-1}$ over timescales of years (the other is N6652B;  \citealt{Paduano21}). Other known bright X-ray transients in globular clusters were either not in outburst during our observations or were not in the original MAVERIC data release. For all sources we assume the cluster distances from \citet{Tremou18}.

\subsection{Radio Observations}

The radio continuum data used in this paper come primarily, though not entirely, from the MAVERIC survey. Observations for the survey were obtained with ATCA or the VLA, depending on the declination of the source. 

Three clusters in this paper have ATCA data: NGC 6441, NGC 6624, and Terzan 6. The ATCA observations are discussed in detail in \citet{Tremou18} and Tudor et al.~(2021, in prep). In brief, for each of these clusters, the observations were made in April 2015 in the extended 6 km configuration. Data were taken in two basebands centered at 5.5 and 9.0 GHz, each with 2 GHz of bandwidth. These data were flagged and calibrated in Miriad \citep{Sault95} and imaged in CASA (Common Astronomy Software Applications; \citealt{McMullin07}), resulting in median synthesized beams of $\sim2$--$3\arcsec$, depending on frequency.

Four of the clusters we study in this work have VLA data: M15, NGC 6652, NGC 6712, and Terzan 6 (which also has ATCA data). M15 was observed in 2011 during an brightening/flaring event of M15 X-2, prior to the official start of MAVERIC but with similar observing frequencies and depth. The M15 data were calibrated and imaged in AIPS (Astronomical Image Processing System; \citealt{Greisen03}) and CASA. Additional data reduction details for M15 can be found in \citet{Strader12}. NGC 6712 was observed as part of the initial MAVERIC VLA program in 2014, with deep C band observations imaged at central frequencies of 5.0 and 7.4 GHz, and calibrated and imaged in AIPS \citep{Shishkovsky20}.

The 2018 Terzan 6 data were obtained as part of the final tranche of MAVERIC VLA observations (program 18A-081, PI: Shishkovsky). These observations were made from 2018 March 25 to June 3 in five separate 1 or 2-hr blocks observed when the VLA was in its most extended A configuration. Data were taken in two 2 GHz basebands with 3-bit receivers, with central frequencies of 5.0 and 7.1 GHz after flagging. These data were calibrated and imaged in AIPS, resulting in a typical synthesized beam of $\sim 1.1\arcsec \times 0.4\arcsec$.

Finally, NGC 6652 was observed in May 2017 as part of a MAVERIC follow-up program to study the transitional millisecond pulsar candidate N6652B with joint {Chandra} and VLA data (program SI0399, PI: Tudor; see \citealt{Paduano21} for the analysis of N6652B). The VLA observations consist of a single $\sim 2.5$ hr block in C configuration at X band, using 3-bit receivers and two 2 GHz basebands centered at 9.0 and 11.0 GHz. The data were flagged, calibrated, and imaged in CASA. The synthesized beam at the average frequency of 10 GHz was $5.3\arcsec \times 2.1\arcsec$.

The observation times and 5.0 GHz flux densities (or upper limits) for all sources with accompanying X-ray measurements are listed in Table \ref{tab:quasisim}, and each source is discussed in detail in Section 3. For the ATCA observations, which are centered at 5.5 GHz rather than 5.0 GHz, we infer the 5.0 GHz density from the 5.5 GHz flux density and the the measured (or assumed) spectral index between 5.5 and 9.0 GHz. We give the details of the calculation for each source in the appropriate subsection.

All radio flux densities were measured by fitting the appropriate point source models to the individual images. Unless otherwise stated, the radio continuum flux density measurements list $1\sigma$ uncertainties (including a minimum 1\% systematic uncertainty), and upper limits are given at the $3\sigma$ level, where $\sigma$ for the upper limits is measured as the root mean square noise in the local region around the source. Radio spectral indices or limits were determined using the Bayesian fitting method described in \citet{Shishkovsky20}. Representative radio continuum images for each source are shown in the Appendix in Figures \ref{fig:fim1} and \ref{fig:fim2}.

\subsection{X-ray Observations}

A key aspect of the MAVERIC survey was obtaining X-ray observations close in time to at least one radio observing block for each cluster, allowing for close-to-simultaneous radio and X-ray measurements for more luminous sources. For clarity, we define ``strictly simultaneous" measurements to be those for which the X-ray and radio data at least partially overlap in time (typically the radio observations last longer). We take ``quasi-simultaneous" observations to be those without overlapping data, but with an offset between the X-ray and radio data of $\lesssim 2$ d.

The simultaneous X-ray measurements for most globular clusters in this paper come from Swift/XRT, excepting M15 and NGC 6652, which are from {Chandra}. {Chandra} can resolve individual sources even in dense clusters. By contrast, individual sources are not typically well-resolved in  Swift/XRT data, but all of the targets in our observations are sufficiently X-ray luminous that they dominate the cluster emission, and hence any Swift/XRT counts can be ascribed to the targets of interest. We analyzed all \xrt\ \citep{Gehrels04, Burrows05} and \chandra\  observations of our sample that were taken close to (or simultaneously with) radio observations of these targets. These observations are listed in Table \ref{tab:quasisim}. The Swift/XRT observations were all made in Photon Counting mode. Unless otherwise stated, all reported X-ray fluxes and luminosities are in the 1--10 keV band, and all X-ray parameter uncertainties are at the 90\% confidence level.

All \xrt\ observations were reduced and analyzed using \textsc{Heasoft}\footnote{\url{https://heasarc.gsfc.nasa.gov/docs/software/lheasoft/}} \citep[version 6.25,][]{HEASARC14}. We reprocessed XRT data using \texttt{xrtpipeline}, and extracted spectra with \texttt{xselect}, following standard procedures.\footnote{ \url{https://www.swift.ac.uk/analysis/xrt/}} 
For observations where the observed source count rate was high enough to cause pileup, we followed the recommended methods to estimate and exclude the piled up region.\footnote{\url{https://www.swift.ac.uk/analysis/xrt/pileup.php}}

We also reduced and analyzed the \chandra\ data fitting our simultaneity criteria using \textsc{CIAO} 4.10 with CalDB 4.7.4 \citep{Fruscione06}. These data include \chandra/ACIS-S data for NGC 6652 (Obs. ID 18987) and \chandra/HRC data for M15 (Obs. ID 13420).

NGC 6652 is the host cluster to XB~1832--330. Observed with a subarray on the ACIS-S S3 chip, XB~1832--330 is likely piled up in this observation. We extracted spectra from the source point spread function and modeled the pileup using the model produced by \citet{Davis01}. For this model we froze the frame time to 0.4s (based on the subarray mode used), the maximum number of photons considered for pileup in a single frame (\texttt{max\_ph}) to 5, grade correction to 1, pileup PSF fraction to 0.95, and left grade migration ($\alpha$) variable. To complement this analysis, we also modeled
the faint, un-piled up readout streak from XB 1832--330 that is also present in this observation. This streak has a relatively low signal-to-noise ratio and so does not offer a precise independent spectral constraint. Nonetheless, analysis of the streak gave a consistent result. This source is discussed in more detail in Section 3.3.

\begin{figure*}[!ht]
  \centering
 \includegraphics[width=1.0\textwidth]{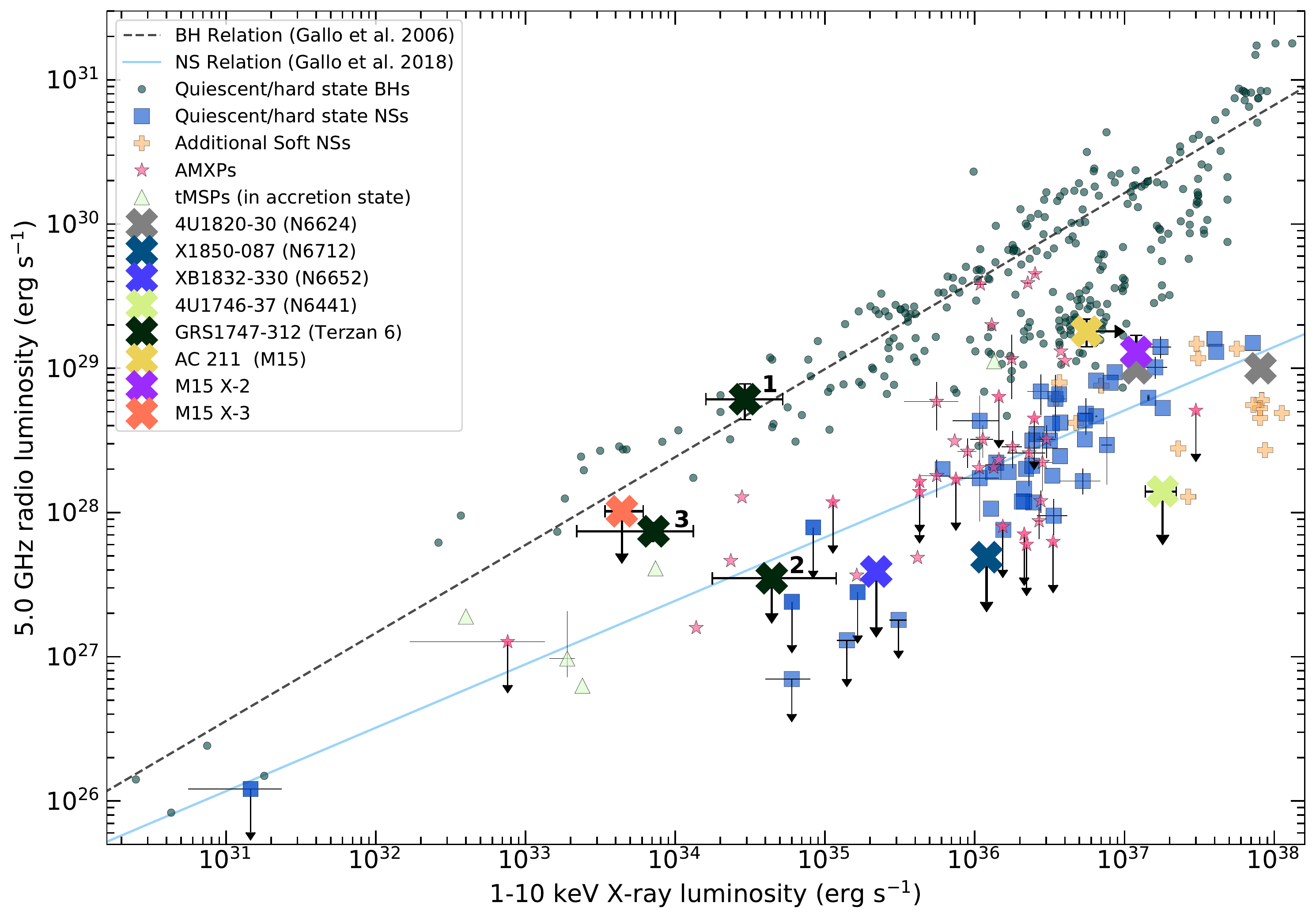}
  \caption{
Radio luminosity (at 5.0 GHz) vs.~X-ray luminosity (1--10 keV) for accreting black hole and neutron star low-mass X-ray binaries, adapted from the compilation of \citet{Bahramian18}. Only sources with simultaneous or quasi-simultaneous data are plotted. The globular cluster sources from this paper are represented as large crosses (the letter ``X"); those with multiple points are sources with more than one quasi-simultaneous data point. For clarity, we have also labeled the points for GRS 1747--312 (in Terzan 6) with numbers, denoting the first (1: 2015 April 18), second (2: 018 March 25), and third (3: 2018 April 30) quasi-simultaneous epochs for this source. The radio flux densities and X-ray fluxes from Table \ref{tab:quasisim} have been converted to luminosities using the cluster distances from \citet{Tremou18}. Dark green circles show quiescent/hard state black holes. Quiescent/hard state neutron stars are shown as blue squares, with additional data points from Aql X-1 \citep{Gusinskaia20a}, and soft sources shown as orange pluses from \citep{Ludlam19}. Pink stars are accreting millisecond X-ray pulsars and light green triangles are transitional millisecond X-ray pulsars. The dashed gray line shows the best-fit relation ($L_R \propto L_X^{0.61}$) for black holes from \citet{Gallo06} and solid blue line represents a proposed correlation ($L_R \propto L_X^{0.44}$) for hard-state neutron stars from \citet{Gallo18}.}
 \label{fig:f1}
  \end{figure*}

For the analysis of the \chandra/HRC data of M15, after standard reprocessing, we estimated count rates and background for each of the three sources, then determined the X-ray fluxes using the model of \citet{White01} (for AC 211 and M15 X-2) or \citet{Arnason15} (for M15 X-3). These data are discussed in more detail in Section 3.6.

We used \textsc{Xspec} \citep{Arnaud96}, with \citet{Verner96} photoelectric cross section and \citet{Wilms00} abundances for spectral analysis. In cases where the number of extracted source events were more than 100, we binned the spectra to have at least 20 counts per bin and used $\chi^2$ statistics for fitting. In other cases, we binned the spectra to have at least 1 count per bin, and used w-statistics as implemented in \textsc{Xspec}.\footnote{\url{https://heasarc.gsfc.nasa.gov/xanadu/xspec/manual/XSappendixStatistics.html}}

The X-ray data used in this paper primarily plays a supporting role to help contextualize the radio observations. Therefore, we do not attempt comprehensive fitting of X-ray models, but instead choose the simplest model(s) that allow us to characterize the flux and spectral state. For some sources, the X-ray flux is low enough that only a simple absorbed power-law or blackbody fit is possible. For other, typically more luminous sources, more complex models such as {\tt Nthcomp} (a thermal comptonized continuum with blackbody seed photons) or composite models are justified and provide good fits. In making these choices we are guided by both the quality of the data and by previous work on these sources. For some of the brighter sources, data from MAXI (all using the Gas Slit Camera; GSC) and Swift/Burst Alert Telescope (BAT) are used to contextualize the Swift/XRT or Chandra data, providing additional constraints on the spectral state or luminosity of the source.

\subsection{New HST Optical Positions} 
To ensure high-fidelity astrometric matching between the X-ray binaries and the radio continuum observations, we determined more precise positions of most of the sources in our sample using new Hubble Space Telescope (HST)/WFC3 observations. For all of M15, NGC 6441, NGC 6624, NGC 6652, and NGC 6712, there are ultraviolet ($F275W$) data that clearly show the sources in question and for which we calculate new positions. The exceptions are AC 211 in M15, which has a precise position from very long baseline interferometry \citep{Kirsten14}, and Terzan 6, which has no ultraviolet data (and indeed the optical counterpart to GRS 1747--312 is unknown). The position of GRS 1747--312 is discussed in more detail in Section 3.5. 

For all HST data, we used a large number of stars detected by HST and in Gaia DR2 \citep{Gaia18} for the ICRS astrometric solutions, finding rms values in the range of 8 to 15 mas per coordinate. We assign conservative uncertainties of 0.02\arcsec\ per coordinate to our final positions, which are listed in Table 1. These are more accurate and precise than existing X-ray positions, and are ideal for use in this paper and for future work. Note that we have made the small necessary adjustments to put the literature positions on ICRS for consistency.

\section{Results for Individual Sources}

\subsection{4U 1820--30 in NGC 6624}

4U 1820--30 is a persistent ultracompact binary with a period of 0.0079 d ($\sim 11.4$ min; \citealt{Stella87}). The short period and high mean X-ray luminosity of the system are consistent with a low-mass ($\sim 0.07 M_{\odot}$) He white dwarf donor \citep{Rappaport87}. The accretor is a neutron star that shows frequent X-ray bursts (e.g., \citealt{NICER19}). While the orbital X-ray variations are of low amplitude ($\lesssim 3$\%), the binary is also observed to have strong X-ray luminosity variations with a superorbital period of $\sim 171 $ d \citep{Priedhorsky84,Zdziarski07}. This superorbital period is typically explained as being due to orbital dynamics in a hierarchical triple (e.g., \citealt{Chou01b,Prodan12}), such that the binary has an as-yet undetected outer companion.

Here we report strictly simultaneous ATCA (radio) and Swift (X-ray) observations of 4U 1820--30 on 2015 April 24--25. We infer a 5.0 GHz flux density of $241.0\pm5.1$ $\mu$Jy based on the measured 5.5 and 9.0 GHz values, and there is an excellent astrometric match (separation of only 16 mas) between the HST optical and ATCA radio positions of 4U 1820--30 \citep{Tremou18}. The radio spectral slope is $\alpha = -0.26\pm0.05$ for an assumed power law with $S_{\nu} \propto \nu^{\alpha}$ where $S_{\nu}$ is the radio flux density at frequency $\nu$ and $\alpha$ is the spectral index (this convention is used throughout this work).

\begin{figure*}[!th]
  \centering
 \includegraphics[width=1.0\textwidth]{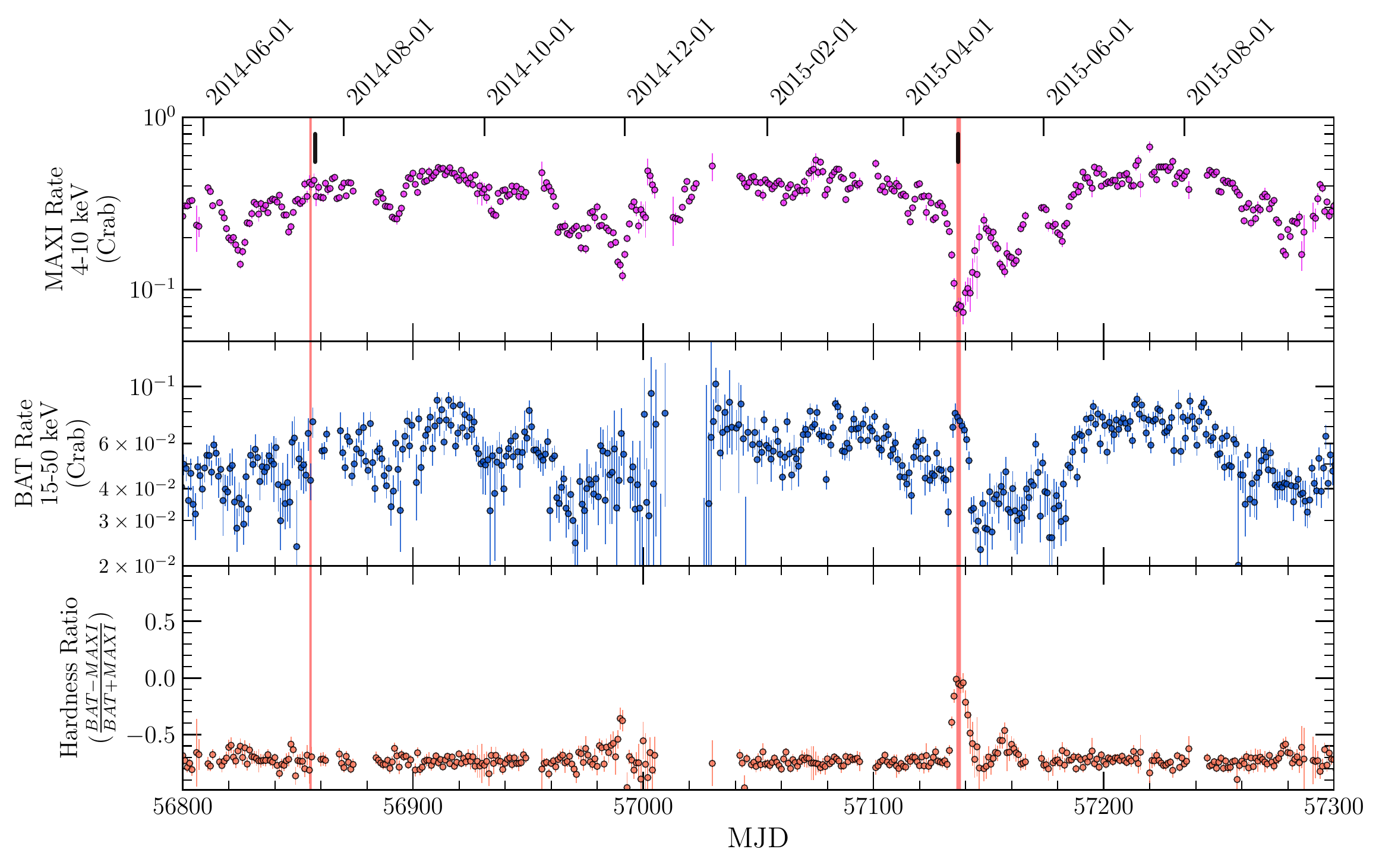}
  \caption{2014 April to 2015 October X-ray light curve of 4U 1820--30 from MAXI (4--10 keV; top) and Swift/BAT (15--50 keV; middle), with hardness ratio in the bottom panel. The epochs of our strictly simultaneous 2015 April Swift/XRT and ATCA data (thick vertical salmon line for radio, short black line for X-ray) as well as the previously published 2014 July quasi-simultaneous radio/X-ray observations (thin vertical salmon line for radio, short black line for X-ray) are also plotted. The new simultaneous data from 2015 April occur during a local minimum in the 
  MAXI flux and a local maximum in the hardness ratio, implying that the system was in the hard island state.}
  \label{fig:4u1820}
 \end{figure*}

The simultaneous Swift/XRT X-ray observations give an unabsorbed 1--10 keV flux of $(1.5\pm0.1) \times 10^{-9}$ erg s$^{-1}$ cm$^{-2}$ when fit by an absorbed {\tt Nthcomp} model. This model is a substantially better fit ($\chi^2$/d.o.f. = 55/40 $\sim 1.4$) than a simple absorbed power law ($\chi^2$/d.o.f. = 118/59 $\sim 2.0$). The respective photon indices for these fits are $\Gamma \sim 1.6$ and $\Gamma \sim 1.3$. In Figure \ref{fig:4u1820} we show the MAXI (4--10 keV) and Swift/BAT (15--50 keV) X-ray light curves for several hundred days around our simultaneous observations. By serendipity, our data were taken nearly perfectly at an unusually low minimum of the superorbital light curve, which corresponds to local maximum of the hardness of the system. Together these measurements strongly suggest that 4U 1820--30 was in the ``island" state of this atoll neutron star LMXB, which is the lowest and hardest state observed for 4U 1820--30 \citep{Bloser00}.

These strictly simultaneous observations place 4U 1820--30 well in the midst of the typical neutron star radio emission observed for sources in the hard/island state at $L_X \sim 10^{37}$ erg s$^{-1}$ (Figure \ref{fig:f1}).

Two previous works also placed 4U 1820--30 on the radio/X-ray correlation, though in the more luminous soft (``banana") state. \citet{diaztrigo2017} presented the 2014 radio continuum detection of 4U 1820--30 with ATCA ($236\pm27 \mu$Jy at 5.5 GHz) and ALMA ($400\pm20 \mu$Jy at 300 GHz; 10 days before the ATCA data). In these 2014 ATCA data, 4U 1820--30 was not detected at 9.0 GHz, with an upper limit of $\lesssim 200 \mu$Jy, implying $\alpha \lesssim -0.34$. We infer a 5.0 GHz flux density of $241\pm28$ $\mu$Jy at this epoch by assuming $\alpha = -0.5$. It is possible that in the soft state 4U 1820--30 sometimes has a steeper radio spectrum than this \citep{Russell21}, but none of our results depend on the precise value assumed. In Swift/XRT observations obtained two days after the ATCA data, 4U 1820--30 is luminous with $L_X \sim 8 \times 10^{37}$ erg s$^{-1}$. These quasi-simultaneous radio/X-ray observations are also plotted in Figure \ref{fig:f1}. The inferred 5.0 GHz flux density was identical to that in the 2015 data, but at an X-ray luminosity a factor of $\sim 6$ higher, where it is in a location typical of soft accreting neutron stars. \citet{Migliari04} presented earlier quasi-simultaneous data in a comparable soft state, finding the radio flux density was about a factor of $\sim 2$ lower than in the \citet{diaztrigo2017} data. We do not include these earlier data in this paper since the system was only significantly detected in an average of seven epochs, and not in any individual epoch \citep{Migliari04}, due to the lower sensitivity of ATCA prior to its backend upgrade in 2010.

\subsection{X1850--087 in NGC\,6712}

\begin{figure*}[!th]
  \centering
 \includegraphics[width=1.0\textwidth]{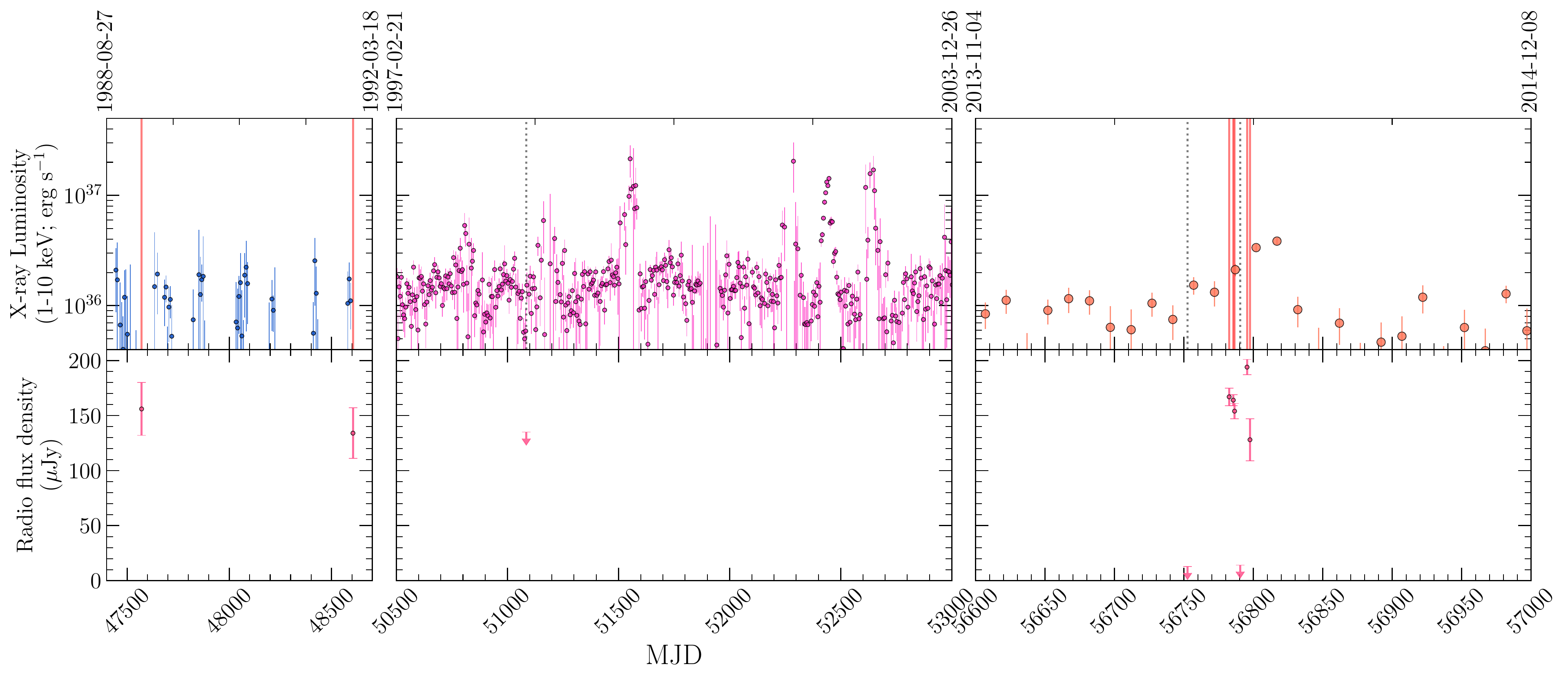}
  \caption{Long-term X-ray light curve of X1850--087 (top) with radio flux densities (bottom). The three separate panels show different time ranges observed with different X-ray instruments: Ginga (blue); {RXTE} (magenta); MAXI (orange), all plotted as 1--10 keV luminosities. Salmon vertical lines correspond to radio detections and grey dashed lines to radio upper limits.}
  \label{fig:f2}
 \end{figure*}

X1850--087 was an early discovery of a persistent luminous X-ray source in a Galactic globular cluster \citep{Sewardetal1976}, with a typical $L_X \sim 10^{36}$ erg s$^{-1}$ (Figure \ref{fig:f2}). The source shows Type I X-ray bursts \citep{Zand19}, proving a neutron star primary. X1850--087 is a likely ultracompact binary with a suspected 0.014 d (20.6 min) orbital period, derived from an HST ultraviolet time series \citep{Homer96}.

\subsubsection{2014 Quasi-simultaneous VLA and Swift Data}

We obtained VLA observations of NGC 6712 over seven epochs, each of duration 1 or 2 hr, in April and May 2014 shown in Table \ref{tab:lrlx}. X1850--087 is strongly variable in the radio over this timespan, with a mixture of tight upper limits and clear detections.

First we discuss our quasi-simultaneous VLA and Swift radio and X-ray observations, with the VLA data obtained on 2014 April 5 and the X-ray data on 2014 April 6. The source is not detected with the VLA at either 5.0 or 7.4 GHz during these observations, with $3\sigma$ upper limits of $< 12.9 \, \mu$Jy and $< 12.3 \, \mu$Jy, respectively. Averaging the subbands gives a $3\sigma$ upper limit  of $< 9.0 \, \mu$Jy at an average frequency of 6.2 GHz.

The quasi-simultaneous Swift/XRT X-ray observations are well-fit by an absorbed power-law with $\Gamma = 2.30\pm0.13$, giving an unabsorbed flux of $(1.6\pm0.1) \times 10^{-10}$ erg s$^{-1}$ cm$^{-2}$. This flux corresponds to an X-ray luminosity of $1.2 \times 10^{36}$ erg s$^{-1}$. We note that a photon index of $\Gamma \sim 2.3$ is atypically soft at this X-ray luminosity \citep{Wijnands15}, perhaps suggesting that the true spectrum is instead a composite that has contributions from both a power law and a thermal disk spectrum. Unfortunately, the quality of the Swift data are not high enough to distinguish between this hypothesis and the simpler single power law. If we try such a two-component fit with the photon index fixed to $\Gamma = 1.7$, both the fit quality and inferred flux are essentially identical, indicating that the presence of a two-component X-ray spectrum is plausible, but unproven. If indeed X1850--087 is in the hard state, then Figure \ref{fig:f1} shows that this source has the most stringent quasi-simultaneous 3$\sigma$ radio continuum upper limit ($L_R \sim 3 \times 10^{28}$ erg s$^{-1}$) for a hard state neutron star LMXB at $L_X \gtrsim 10^{36}$ erg s$^{-1}$. 

While X1850--087 is only marginally detected in daily MAXI (4--10 keV) and Swift/BAT (15--50 keV) observations in this time frame, binned light curves do show detections (Figure \ref{fig:panel3}). These light curves show that there is no meaningful variability in the flux or hardness of X1850--087 in the data surrounding the 2014 April 5--6 radio and X-ray observations, suggesting the quasi-simultaneous Swift/XRT data are representative.

We also delve into the 2014 April 5 radio non-detection by imaging the data in individual 10-min scans, averaging together both basebands in the $uv$ plane. The source was undetected in all the individual scans, with a typical per-scan rms of about 9 $\mu$Jy (and hence a per-scan $3\sigma$ limit of $\lesssim 27 \mu$Jy). We conclude that the limit in the entire 2 hr block is representative and that the binary is indeed not detected in these radio continuum data at this time.

\subsubsection{2014 May Data: Detections and Variability}

In contrast to the strong radio upper limit in 2014 April, X1850--087 is clearly detected in several other radio observations in 2014 May at $> 10\sigma$ significance (Table \ref{tab:x1850}). Of the six VLA epochs obtained during May 2014, in five of these X1850--087 has a flux density $>100$ $\mu$Jy. In one epoch (2014 May 13), X1850--087 is again not detected (with a formal $3\sigma$ upper limit of $< 9.5 \mu$Jy at a combined average frequency of 6.2 GHz), but is well-detected in bracketing observations on May 9 and May 18. This demonstrates that the radio flux density of X1850--087 is changing by at least a factor of $\sim 20$ in just a few days.

\begin{figure}[!t]
  \centering
 \includegraphics[width=0.5\textwidth]{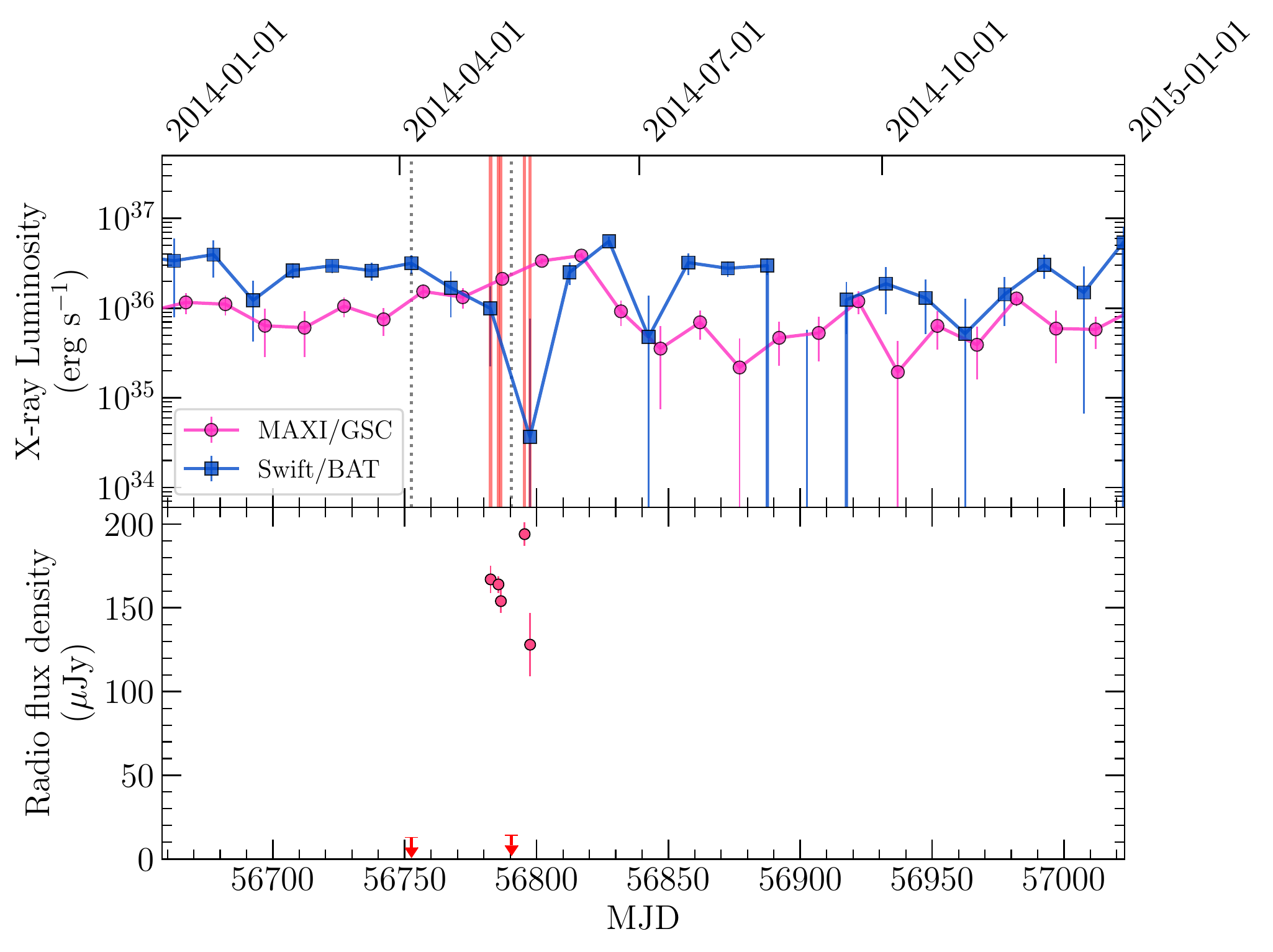}
  \caption{2014 X-ray light curve of X1850--087 (top) with radio flux densities (bottom). Both MAXI (4--10 keV) and Swift/BAT (15--50 keV) X-ray light curves are shown. As in Figure \ref{fig:f2}, salmon vertical lines correspond to radio detections and grey dashed lines correspond to radio upper limits.}
  \label{fig:panel3}
 \end{figure}

While there is no Swift/XRT X-ray coverage of these 2014 May epochs, the binned MAXI and Swift/BAT light curves indicate that an X-ray brightening and a transition to a softer state occurred around the time of these 2014 May VLA observations (Figure \ref{fig:panel3}). Hence, one interpretation of the overall higher radio flux density from 2014 May 5--9 and May 18--20 is that it is associated with this transition, perhaps due to the launching of discrete ejecta. A challenge to this simple interpretation is the non-detection on May 13, which would require a separate transition to the baseline fainter harder state (or perhaps an intermediate hard state) between May 9 and 13 and a re-flaring to a softer higher state by May 18. As we lack daily Swift/XRT data during this period we cannot definitively rule out this possibility, but it is not supported by the binned MAXI and Swift/BAT light curves, and would also represent phenomenology not previously observed in the {Rossi X-ray Timing Explorer (RXTE)} light curve for the source (see next subsection).

Absent this fast flaring behavior, we know of no straightforward explanation for such dramatic variability on these timescales in an ultracompact system, which are much longer than the 0.014 d (20.6 min) orbital period but much shorter than, for example, the $\gtrsim 100$ d superorbital periods observed in some other ultracompact systems such as 4U 1820--30 (Section 3.1).

In the May 13 radio data for X1850--087, as in the April 5 data, the source is not detected in 10-min individual scans during the 2 hour observation block. This is consistent with the idea that any orbital variations---which ought to be averaged out over individual blocks that represent $\sim 6$ orbital periods---are not the primary cause of the variability of X1850--087.

We also imaged the individual scans for the observing blocks adjacent to the May 13 block in which the source was not detected. Each of these is a 1-hr block with five 8-min scans on source. On May 9, there is no evidence for significant variation among the scans. On May 18, X1850--087 is brighter in the last two scans than the first three at a formal level of about $3.5\sigma$ (5.0 GHz flux density of $235\pm17 \mu$Jy in the last two, compared to $168\pm9 \mu$Jy in the first three). Given its marginal significance, we do not attach too much import to the change, but if it were a real increase in the flux density, it could be short-term variability due to a recent jet ejection event.

We report spectral indices in Table \ref{tab:x1850} for our 2014 May observations of X1850--087. The spectral index ranged from inverted (brighter at higher frequencies) to flat, varying on short timescales: between May 5 and May 8 the source changed from an inverted spectral index ($\alpha = +0.50\pm0.15$) to a flatter value ($\alpha = -0.13\pm0.11$), then quickly back to inverted ($\alpha = 0.46\pm0.13$) on May 9, just one day later. The detection on May 18 was likewise flat, with $\alpha = +0.06\pm0.13$. We also calculated the per-scan spectral indices on May 18,
finding a mean value of $\alpha=-0.04\pm0.20$ for the three early scans and $\alpha =+0.30\pm0.16$ for the two final scans. This could be consistent with a transition to more inverted emission during the putative flux density increase at the end of the May 18 observing block, though the evidence for a spectral index change is not formally significant. We discuss possible interpretations of the spectral index variations between different epochs in Section 4.2.3.

\subsubsection{Other Radio and X-ray Data}
To improve the time baseline in the radio, we also make use of archival VLA data for X1850--087, reducing all of the C band data taken in the high-resolution (A or B) VLA configurations. This resulted in three additional epochs, all at 4.9 GHz, spanning 1989--1998 (Table \ref{tab:x1850}). In 1989 February and 1991 December, X1850--087 was clearly detected, with flux densities consistent with our 2014 detections. In 1998 September, the source was not detected, but the 3$\sigma$ upper limit ($<135 \mu$Jy) is not particularly constraining as to whether order-of-magnitude variability was present in these older data.

Some previous papers on the 1989 February radio detection have claimed that X1850--087 is spatially resolved in these data \citep{Lehtoetal1990,Machinetal1990}, which if true would have important implications for its interpretation. In our re-reduction of these data, given the moderate signal-to-noise of the source, we find that either a point source or extended morphology offer reasonable fits. In our substantially deeper 2014 detections while the VLA was in its highest resolution A configuration, we find that in all epochs the morphology of X1850--087 is consistent with a point source.

There is at least partial X-ray data at these earlier epochs to contextualize these radio detections and limits. A Ginga/ASM (1--20 keV) light curve is available from 1987 March to 1991 October, and while it does not have spectral information, this X-ray light curve shows only modest variability (Figure \ref{fig:f2}). Two epochs of Ginga/LAC data in this time interval, from 1989 April and 1990 Sep, suggest an unabsorbed 1--10 keV luminosity of $L_X \sim 1.8 \times 10^{36}$ erg s$^{-1}$ \citep{Kitamotoetal1992}, consistent with the Swift/XRT data from 2014. 

{RXTE}/ASM (1--10 keV) offers a well-sampled light curve of X1850--087 from 1995 to 2012, with evidence for a number of distinct episodes of brightening on typical timescales of weeks to months, as well as lower-level variability. In the vicinity of the 1998 September radio upper limit, the source has an X-ray flux typical of the long-term value \citep{Cartwright13}.

The most notable feature of the long-term X-ray light curve of X1850--087 is that it spends a relatively short amount of time in obvious bright states. As a rough estimate, we use the {RXTE}/ASM light curve, which has the best sampling and decent sensitivity of the long-term light curves. If we make a somewhat arbitrary designation that the source is in a bright state if the X-ray luminosity is $\gtrsim 3 \times 10^{36}$ erg s$^{-1}$, then it spends only $\sim 6\%$ in such bright states (see also \citealt{Cartwright13}). This percentage increases to $\sim 15\%$ if a more conservative luminosity of $\gtrsim 2 \times 10^{36}$ is used, though Figure \ref{fig:f2} shows that this value would likely also encompass periods of typical persistent activity outside of a bright state. Hence the long-term X-ray light curve shows that it is unlikely that two random radio continuum observations in 1989 and 1991 would have been taken during this uncommon flaring/bright state purely by chance. 

Overall, we conclude that a straightforward flaring/state transition model for the dramatic radio variability of X1850--087 cannot be firmly ruled out. However, such an scenario also does not readily explain the data.

\subsection{XB1832--330 in NGC\,6652}

XB1832--330 is a persistent X-ray source first associated with the globular cluster NGC 6652 in 1990 \citep{Predehl91} and with evidence for Type I X-ray bursts \citep{intZand98,Mukai00}. Based on its X-ray spectrum and the lack of a bright optical counterpart, it was long argued to be an ultracompact LMXB (e.g., \citealt{Heinke01,Parmar01}). However, \citet{Engel12} used Gemini optical photometry to argue for an orbital period of $0.0895\pm0.0001$ d (2.15 hr), suggesting instead a low-mass main sequence donor.

\begin{figure*}[!th]
  \centering
 \includegraphics[width=1.0\textwidth]{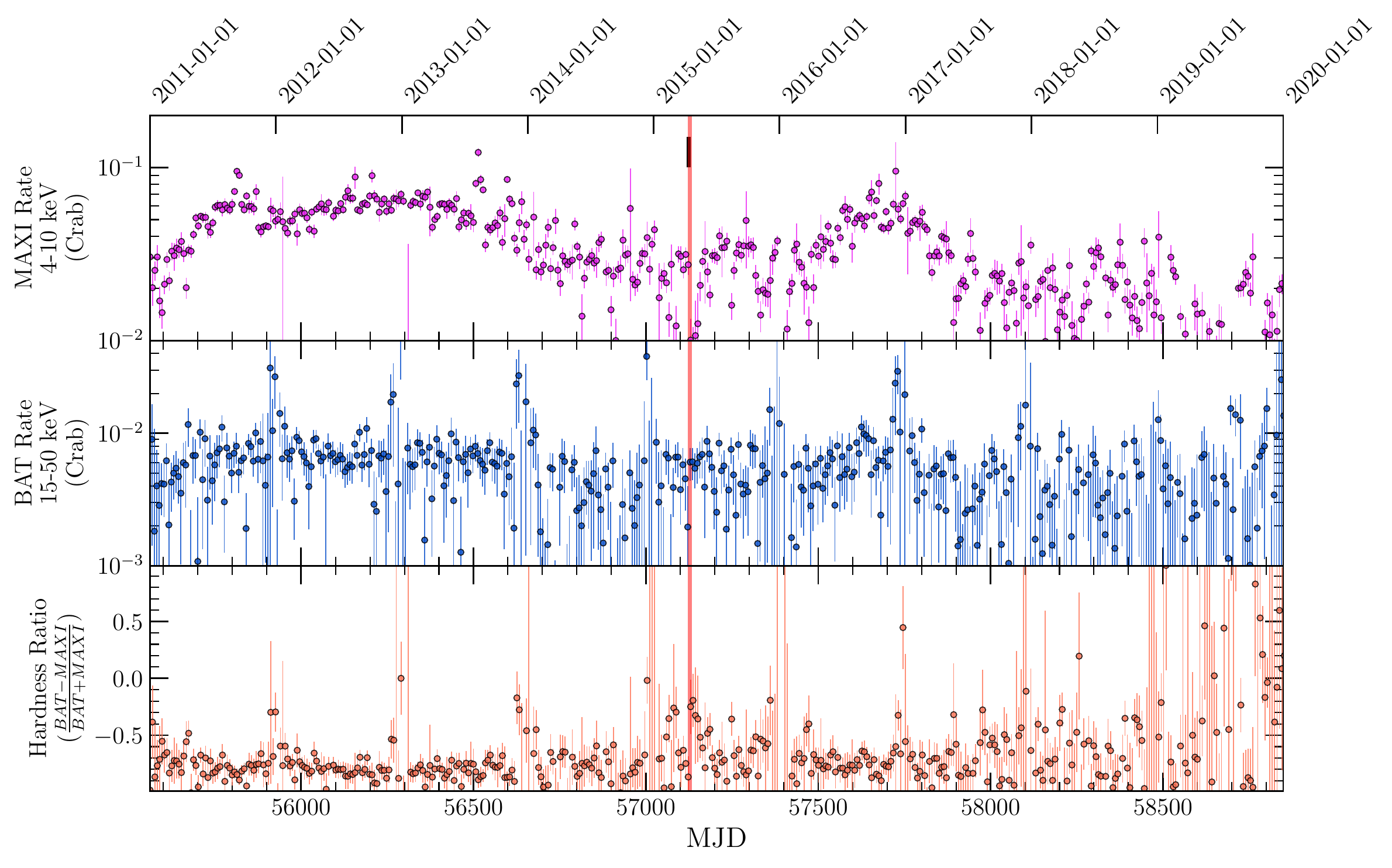}
  \caption{Long-term X-ray light curve of 4U 1746--37 from MAXI (4--10 keV; top) and Swift/BAT (15--50 keV; middle), with hardness ratio in the bottom panel. The epoch of our quasi-simultaneous 2015 April Swift/XRT and ATCA data (thick vertical salmon line for radio, short black line for X-ray) are marked.}
  \label{fig:4u1746}
 \end{figure*} 

Here we report a $3\sigma$ VLA radio upper limit of $< 6.6 \mu$Jy at 10 GHz on 2017 May 22. While we have strictly simultaneous {Chandra}/ACIS-S observations, these are piled up and had to be specially fit (as described in Section 2.3). An absorbed {\tt Nthcomp} model provides a good fit ($\chi^2$/dof = 199/186 $\sim 1.07$) with an unabsorbed X-ray flux of $(1.9\pm0.1) \times 10^{-11}$ erg s$^{-1}$ cm$^{-2}$. The system was in the hard state with $\Gamma = 1.62\pm0.14$.  

The X-ray flux reported above corresponds to a luminosity of $L_X \sim 2\times10^{35}$ erg s$^{-1}$. This value is somewhat fainter than that reported for the 2011 {Chandra} observations of \citet{Stacey12}, and much lower than the $L_X \gtrsim 10^{36}$ erg s$^{-1}$ that the source showed when observed from 1990 through 2010 (e.g., \citealt{Predehl91,Parmar01,Sidoli08,Cartwright13}). {RXTE}/PCA Bulge Scan data\footnote{\url{https://asd.gsfc.nasa.gov/Craig.Markwardt//galscan/html/R_1832-330.html}} suggests that a transition to a fainter state occurred beginning around 2011 February. There is some evidence from {HEAO-1} observations that in the late 1970s the luminosity was closer to $\sim 10^{35}$ erg s$^{-1}$ \citep{Hertz85}, implying that similar persistent luminosity changes have occurred in the past.

Our deep radio upper limit at $L_X \sim 2\times10^{35}$ erg s$^{-1}$ puts XB1832--330 among the sources with the most constraining simultaneous radio data at these ``low" persistent emission levels. 

\subsection{4U 1746--37 in NGC\,6441}

4U 1746--37 is a persistent atoll X-ray source in NGC 6441, with an orbital period of 0.215 d (5.14 hr) as inferred from {RXTE} X-ray dips \citep{Balucinskachurchetal2004}. The radio data implicitly average over several orbital periods, though the dips have maximal X-ray amplitudes of $\sim 25\%$. The dipping nature of the source suggests an inclination $\gtrsim 60^{\circ}$, possibly as high as $\sim 75-80^{\circ}$.

We report a $3\sigma$ ATCA 5.5 GHz radio upper limit of $< 13.7 \mu$Jy on 2015 April 14/15. Quasi-simultaneous Swift X-ray observations on 2015 April 13 are best-fit by a composite {\tt Nthcomp}+{\tt diskbb} model, yielding an unabsorbed X-ray flux of $(8.6\pm2.0) \times 10^{-10}$ erg s$^{-1}$ cm$^{-2}$. This spectrum and flux are consistent with that of the soft/banana-like state in which 4U 1746--37 spends most of its time (e.g., \citealt{Jonker00,Munoz14}). The Swift/BAT and MAXI X-ray light curves
around the time of the X-ray and radio observations (Figure \ref{fig:4u1746}) are noisy, requiring coarse 10-day binning, and somewhat difficult to interpret. The single MAXI bin closest in time to the radio observation has a low flux, and combined with the Swift/BAT detection in the corresponding bin, this could suggest the system was in an intermediate hard state. On the other hand, the Swift/BAT flux was no higher than any surrounding bin, and on average the source is in a relatively soft state around this time. Hence we proceed assuming that the system is indeed in the soft/banana state at the time of our radio data.

While 4U 1746--37 shows both bursts and flares, the quasi-simultaneous 2015 April Swift/XRT flux corresponds to an X-ray luminosity of $\sim 2 \times 10^{37}$ erg s$^{-1}$, consistent with a typical non-bursting value for the source (e.g., \citealt{Christian1997}). 

The non-detection of this source in radio continuum emission is notable: it has one of the most constraining radio upper limits for a soft state neutron star at $L_X > 10^{37}$ erg s$^{-1}$ (Figure \ref{fig:f1}), suggesting either relatively faint radio emission from 4U 1746--37 in this soft state or perhaps the quenching of its jet.

\subsection{GRS 1747--312 in Terzan 6}

\subsubsection{Radio and X-ray Measurements}

GRS 1747--312 is a transient LMXB first observed in 1990 by Granat \citep{Pavlinsky94}. The compact object is a neutron star that shows Type I X-ray bursts \citep{intZand03b}.  Observations of X-ray eclipses have allowed the determination of a precise orbital period of 12.360 hr \citep{intZand03a}. GRS 1747--312 has unusually frequent outbursts, clustering with a recurrence time around 4.5 months, though some outbursts occur much more quickly or slowly than this typical value. While the optical companion is unknown---due primarily to the high optical extinction toward Terzan 6---the relatively long orbital period and fast recurrence time suggest a somewhat high mass transfer rate from a subgiant donor \citep{intZand03a}.

Here we report three quasi-simultaneous (within $\sim 1$ d) epochs of radio and X-ray observations of GRS 1747--312. The ephemerides of the X-ray eclipses are known with sufficient precision \citep{intZand03a} that we can determine with certainty that none of the X-ray data discussed below occur during eclipses. For one of the five VLA epochs (2018 June 3), and for the ATCA observations, there is a partial overlap with the predicted eclipse times. The radio emission at these frequencies (5/7 GHz for VLA; 5.5/9 GHz for ATCA) is unlikely to be eclipsed even during X-ray eclipses \citep{Maccarone20}, and in any case only a small amount---about 13\% of the 2018 June 3 VLA epoch and $\lesssim 4\%$ of the ATCA data---was taken during a predicted eclipse, so does not affect our results or conclusions.

In ATCA observations that began 2015 April 17 (labeled as 1 in Figure \ref{fig:f1}), the binary is well-detected with 5.5 and 9.0 GHz flux densities of $213.3\pm5.1 \, \mu$Jy and $172.4\pm5.2 \, \mu$Jy, respectively, implying a radio spectral index $\alpha =$ --$0.43\pm0.08$. We use this well-measured spectral slope to infer a 5.0 GHz flux density of $221.1\pm6.4 \, \mu$Jy. The data quality of all three Swift/XRT observations discussed in this section (including the quasi-simultaneous one obtained on 2015 April 16) is relatively low, and fitting an absorbed power-law to these X-ray data gives a strong degeneracy between $\Gamma$ and $N_H$. The presence of variable dips for this source means that a constant $N_H$ cannot necessarily be assumed. Therefore we have taken the approach of fitting two models to each of the Swift/XRT observations: one with fixed $\Gamma$ and free $N_H$, and the other with both $\Gamma$ and $N_H$ fixed, the latter to the estimated foreground of $1.4 \times 10^{22}$ cm$^{-2}$. Motivated by the typical photon indices for low-mass X-ray binaries at these X-ray luminosities and the spectral fitting results of \citep{Vats18} for the source in quiescence, for the models with $\Gamma$ fixed, we assume $\Gamma = 1.7$. 

For the 2015 April 16 Swift/XRT data, the free $N_H$ fit has $N_H = 7.8^{+6.4}_{-4.0} \times 10^{22}$ cm$^{-2}$ and an unabsorbed X-ray flux of $5.3^{+4.0}_{-2.4} \times 10^{-12}$ erg s$^{-1}$ cm$^{-2}$, corresponding to $L_X = 2.9^{+2.2}_{-1.3} \times 10^{34}$ erg s$^{-1}$ at the assumed distance of 6.8 kpc. For the fit with both $N_H$ and $\Gamma$ fixed, we find an unabsorbed flux of $2.1^{+1.0}_{-0.8} \times 10^{-12}$ erg s$^{-1}$ cm$^{-2}$ ($L_X = 1.2^{+0.5}_{-0.4} \times 10^{34}$ erg s$^{-1}$). These two flux estimates are consistent within the uncertainties; we plot the former (free $N_H$) fit in Figure \ref{fig:f1}, but none of our conclusions depend on this choice.

At this first quasi-simultaneous epoch, GRS 1747--312 is the most radio-luminous neutron star LMXB ever observed at X-ray luminosities $< 10^{35}$ erg s$^{-1}$, with a 5.0 GHz radio luminosity of $(6.1\pm0.2) \times 10^{28}$ erg s$^{-1}$. It sits near the upper envelope of the well-defined black hole radio/X-ray correlation (Figure \ref{fig:f1}).

Two VLA epochs in 2018 were obtained quasi-simultaneously with additional Swift/XRT data. On 2018 March 25 (labeled as 2 in Figure \ref{fig:f1}), GRS 1747--312 is not detected with the VLA at either 5.0 or 7.1 GHz, with $3\sigma$ upper limits of $< 12.9 \, \mu$Jy and $< 11.7 \, \mu$Jy respectively. We fit the Swift/XRT observations on 2018 March 26 using the two models discussed above: $\Gamma = 1.7$ and $N_H$ free, or $\Gamma = 1.7$ and $N_H = 1.4 \times 10^{22}$ cm$^{-2}$. For the former model we find $N_H = 3.6^{+4.2}_{-2.1} \times 10^{23}$ cm$^{-2}$ and an unabsorbed X-ray flux of $8.0^{+13.6}_{-4.8} \times 10^{-12}$ erg s$^{-1}$ cm$^{-2}$, equivalent to $L_X = 4.4^{+7.5}_{-2.6} \times 10^{34}$ erg$^{-1}$. For the fixed $N_H$ model, the unabsorbed flux is $1.0^{+0.6}_{-0.4} \times 10^{-12}$ erg s$^{-1}$ ($L_X = 5.5^{+3.3}_{-2.4} \times 10^{33}$ erg$^{-1}$). For this observation, the X-ray fluxes from the two models are not quite consistent within the uncertainties. We conservatively plot the free $N_H$ fit in Figure \ref{fig:f1}, but again none of our conclusions depend on the precise X-ray luminosity plotted for this radio upper limit.

GRS 1747--312 is again detected in the radio on 2018 April 30 (labeled as 3 in Figure \ref{fig:f1}) with 5.0 and 7.1 GHz flux densities of $26.9\pm4.3 \, \mu$Jy and $32.1\pm4.1 \, \mu$Jy respectively, and a measured spectral index of $\alpha=+0.51\pm0.57$. In quasi-simultaneous Swift/XRT data from 2018 May 1, it is marginally detected, with an unabsorbed X-ray flux from the free $N_H$ model of $0.9^{+1.1}_{-0.6} \times 10^{-12}$ erg s$^{-1}$ cm$^{-2}$ ($N_H = 0.5^{+1.8}_{-0.5} \times 10^{22}$ cm$^{-2}$) and a fixed $N_H$ flux of  $(1.3^{+1.2}_{-0.8}) \times 10^{-12}$ erg s$^{-1}$ cm$^{-2}$. Since the former fit yields an $N_H$ below the expected foreground, for this epoch we use the fixed $N_H$ result in Figure \ref{fig:f1}, again emphasizing that this choice does not affect any of our conclusions. This gives $L_X = 7.1^{+6.4}_{-4.4} \times 10^{33}$ erg s$^{-1}$.

The VLA non-detection in March 2018 is more typical of the upper limits or detections for neutron star LXMBs at $L_X$ of $\sim 5\times 10^{33}$ to $5\times 10^{34}$ erg s$^{-1}$. However, the VLA detection about a month later (2018 April 30), at a fainter X-ray and radio luminosity than the bright 2015 detection, again approaches the regime more typically occupied by black hole LMXBs. In subsequent VLA epochs, taken from 2018 May 21 to June 3, the source is well-detected at a comparable radio luminosity to the first VLA detection (Table \ref{table:grstable}), also showing modest (less than a factor of $\sim 2$) variability. We have no simultaneous X-ray data at these latter radio epochs. We stack the four detected VLA epochs in the $uv$ plane to get our best estimate of the VLA source position (Table \ref{table:grstable}). The spectral index in the stack is $\alpha=-0.51\pm0.28$, entirely consistent with the ATCA value. This suggest a similar origin for the radio continuum emission in both datasets, despite a factor of $\sim 6$ drop in radio luminosity.

As we discuss below for AC 211 in M15 (Section 3.6.1), given the high orbital inclination of GRS 1747--312, it is possible that its unabsorbed X-ray luminosity is being underestimated, which might make GRS 1747--312 somewhat less of an outlier in the radio/X-ray correlation. Mitigating against this possibility is that for GRS 1747--312 the central accretor is visible (unlike the case for AC 211), implying less ``extra" obscuration for GRS 1747--312, if any.

\subsubsection{Are We Sure We are Detecting GRS 1747--312?}

It is worth considering whether we are sure that the radio and X-ray fluxes indeed reflect those of GRS 1747--312. The source was not in X-ray outburst during any of our observations, and the X-ray data were taken outside of eclipse.

While no other luminous X-ray sources are definitively known in Terzan 6, there are some puzzling observations that could be consistent with the existence of sources other than GRS 1747--312. In a 2009 {Suzaku} observation, \citet{Saji16} observe a single source with $L_X \sim 6 \times 10^{34}$ erg s$^{-1}$, but see no evidence for an eclipse at the predicted time for GRS 1747--312, leading them to argue that this X-ray emission might arise from a different source. 

Only a single X-ray source is evident in our Swift/XRT images of Terzan 6, and its position is consistent with the {Chandra}/HRC position of GRS 1747--312 \citep{intZand03b}. But the Swift position is at best only constrained at the level of $\sim 2$--3\arcsec, so this association is not definitive in the dense environment of a globular cluster.

The position of the counterpart radio source in our ATCA and VLA observations is much better constrained. Since this source shows radio variability, has a radio spectral slope consistent with a LMXB, and already has a relatively high ratio of radio to X-ray luminosity, it seems fair to assert that this radio emission indeed arises from the same source as in the quasi-simultaneous Swift/XRT observations. The ATCA and VLA positions of the source can be found in Table \ref{table:grstable}, where we have conservatively assumed the $1\sigma$ uncertainties to be 10\% of the beam. The ATCA and VLA positions are entirely consistent with each other, and the VLA position is about 0.7\arcsec\ from the best-fit {Chandra}/HRC position\footnote{This is not due to the proper motion of Terzan 6, which only implies a change of about 0.1\arcsec\ per coordinate between the X-ray epoch 2000 and and radio epochs of 2015--2018 \citep {Vasiliev19}.}. This X-ray position has a formal 95\% uncertainty of 0.4\arcsec\ \citep{intZand03b} based on an astrometric solution from three X-ray sources in the field. Hence, formally our radio continuum position is inconsistent with the {Chandra} one. Since absolute astrometry with {Chandra} is challenging, it seems plausible that the uncertainty is slightly underestimated, which could reconcile the positions.

A piece of evidence in favor of this argument is that neither the {Chandra}/HRC position of GRS 1747--312 nor our radio position are right at the center of the cluster: rather, they are northwest of the center in an HST/WFC3 F110W image (corrected to ICRS using Gaia DR2 stars), perhaps $\sim 1.4$--2\arcsec\ from the center and approaching the edge of the core. While it is possible that there are two unrelated X-ray binaries only $0.7\arcsec$ from each other but offset from the cluster center, an unrelated binary could plausibly be located anywhere in the core.

Finally, an {XMM} observation of Terzan 6 taken when GRS 1747--312 was not in outburst showed evidence for a single X-ray source with $L_X \sim 10^{33}$ erg s$^{-1}$ (assuming our distance of 6.8 kpc). A spectral fit to this source required a two-component model with both a neutron star atmosphere and a hard power law, consistent with ongoing low-level accretion onto a neutron star \citep{Vats18}. This X-ray luminosity is well below that of our Swift/XRT observations, which range from $L_X \sim 7\times 10^{33}$ to $4\times 10^{34}$ erg s$^{-1}$. Hence, to explain our Swift/XRT data, a non-GRS 1747--312 source would need to have a typical luminosity (in both 2015 and 2018) around $L_X \sim 10^{34}$ erg s$^{-1}$. Being in neither quiescence nor full outburst, such luminosities are quite unusual for LMXBs, but are not unexpected for GRS 1747--312 given its very frequent outbursts. A hypothetical non-GRS 1747--312 source would also need to be X-ray and radio variable, as well as being very close to the position of GRS 1747--312. This verges on implausibility, and the most parsimonious explanation---pending, of course, additional X-ray and radio data---is that the only X-ray binary clearly detected thus far in Terzan 6 is indeed GRS 1747--312.

\subsection{Sources in M15: AC 211, X-2, and X-3}

Here we discuss the persistent sources AC 211 and M15 X-2 together with the very faint X-ray transient M15 X-3, since all the analysis comes from identical datasets. All the radio flux densities for our VLA data can be found in Table \ref{tab:m15}.

\subsubsection{AC 211}

AC 211 (X2127+119) is an LMXB with a 17.1 hr period \citep{Ilovaisky93}. It is one of the most famous accretion disk corona sources, where an edge-on inclination leads to the outer disk blocking the view of the central accretion flow, and only scattered emission is observed. The compact object is generally assumed to be a neutron star but due to the obscuration this has not been confirmed, since the normal identification tools (e.g., Type I X-ray bursts) cannot be used. AC 211 is a variable source that shows eclipses and flares (e.g., \citealt{Hannikainen05}), with milliarcsecond-scale radio outflows or ejection events observed during outbursts \citep{Kirsten14}.

AC 211 is detected in all five epochs of radio continuum imaging in 2011 (see Table \ref{tab:m15}) and is strongly variable, with its 5.0 GHz flux density ranging from 271 $\mu$Jy to 794 $\mu$Jy on timescales as short as a few days. Radio eclipses, while a priori unlikely at these frequencies \citep{Maccarone20}, are one possible explanation for this variability. Unfortunately, the existing ephemerides \citep{Ioannou03} extrapolated to the epoch of observation are not of sufficient precision, with a $1\sigma$ uncertainty of 0.23 d. While the absolute phase cannot be ascertained, the better-determined relative phases of the three low flux density epochs ($< 300 \mu$Jy) also do not line up, with a spread of $\phi = 0.35$ in the mid-exposure times. Hence, if radio eclipses are the main explanation for the variability, they would need to last for longer than a third of the orbit, which is quite unlikely. Improved ephemerides would allow a test of this scenario.

In the single quasi-simultaneous epoch, we find an unabsorbed {Chandra}/HRC X-ray flux of $(4.5\pm0.1) \times 10^{-10}$ erg s$^{-1}$ cm$^{-2}$ (using the parameters of the power-law model of \citealt{White01}), equivalent to $(5.7\pm0.1) \times 10^{36}$ erg s$^{-1}$, and a 5.0 GHz radio flux density of $292.0\pm6.3$ $\mu$Jy. This places it among the more radio-luminous neutron star LMXBs. At the higher, non-simultaneous value of 794 $\mu$Jy, the radio luminosity is a factor of 2.7 higher and hence more typical of stellar-mass black holes than neutron stars, though this higher radio flux density could be associated with a flare rather than steady emission. 

It is worth noting that the X-ray luminosity measurement is almost certainly an underestimate, given the obscuration that prevents us from seeing the central source; the true $L_X$ could be at least an order of magnitude higher than we estimate. We indicate this in Figure \ref{fig:f1} by plotting the $L_X$ measurement as a lower limit. Nonetheless, even at a much higher value of $L_X$, AC 211 would still be the among the most radio-luminous accreting neutron stars observed, especially in the latter non-simultaneous epoch.

One possible explanation is that AC 211 has an intrinsic $L_X$ that is around two orders of magnitude higher than observed and is a ``Z source" accreting near the Eddington luminosity. The nearest and best-studied member of this class, Sco X-1, shows rapid variations in radio flux density and spectral index that are somewhat similar to those that we observe in AC 211 \citep{Hjellming90,Fomalont01}.

Another, speculative possibility is that AC 211 contains a black hole rather than a neutron star, which could potentially explain its radio loud nature. This is statistically unlikely, and evidence from optical spectroscopy tends to favor a neutron star primary  \citep{vanZyl04a,vanZyl04b}, but this prospect cannot be definitively ruled out.

\subsubsection{M15 X-2}

\citet{White01} used {Chandra} data to solve the mystery of why the obscured source AC 211 occasionally showed X-ray bursts that ought not to have been visible: these bursts are associated with a very nearby (separated by $< 3\arcsec$) luminous LMXB, M15 X-2. \citet{White01} found an ultraviolet-bright counterpart to M15 X-2 with a 23-min periodicity identified with the orbital period \citep{Dieball05}, showing that M15 X-2 is an ultracompact X-ray binary---one of only 4 confirmed persistent ultracompact systems known among Galactic globular clusters.

M15 X-2 had bright flaring events in both 2011 and 2013 \citep{Sivakoff11,Pooley13}, with X-ray bursts detected during both events \citep{Negoro16}. The 2011 flare is the one covered by the same five epochs of VLA data as for AC 211. M15 X-2 was strongly variable during this event (see Table \ref{tab:m15}), dropping from a 5.0 GHz flux density $\sim 150$ to $64 \mu$Jy from 2011 May 22 to May 26, then back up to $205.0\pm6.3 \mu$Jy on 2011 May 30. When re-visited in late August 2011, only upper limits were found, equivalent to a $3\sigma$ upper limit of $< 12.7 \mu$Jy at 5.0 GHz if both non-detection epochs are combined.

This radio time series clearly shows that the data from 2011 May 30 reflect an elevated level of radio emission during the brightening event. These data share the same quasi-simultaneous {Chandra}/HRC dataset used for AC 211. At this epoch the unabsorbed {Chandra} flux was $(9.46\pm0.03) \times 10^{-10}$ erg s$^{-1}$ cm$^{-2}$ (using the power-law model of \citealt{White01}), equivalent to $L_X = 1.2 \times 10^{37}$ erg s$^{-1}$. The radio/X-ray luminosity ratio on 2011 May 30 for M15 X-2 is high, close to the upper envelope of other accreting neutron star binaries observed at this X-ray luminosity.

We note that since these {Chandra} data were obtained with HRC, no spectral information was available to assess the state of M15 X-2 at this epoch. Instead, we analyze a Swift/XRT observation from 2011 May 30 which is strictly simultaneous with this VLA epoch. The Swift data has lower spatial resolution than {Chandra} and hence also includes AC 211. Since the count rate of AC 211 in the {Chandra}/HRC data is about 10\% of that of M15 X-2, it makes a small but non-negligible contribution to the Swift/XRT spectrum. Using this relative count rate and the power-law spectral model of \citet{White01} for AC 211, we fit a combination of AC 211 and M15 X-2 to the Swift spectrum. We find that M15 X-2 is well-fit by a composite disk ({\tt diskbb}) + power-law model ($\Gamma = 1.6\pm0.1)$ with an unabsorbed X-ray flux of $(1.04\pm0.04) \times 10^{-9}$ erg s$^{-1}$ ($L_X = 1.3\times10^{37}$ erg s$^{-1}$). About three quarters of the flux in this band is contributed by the power-law, suggesting that M15 X-2 is closer to the hard state than the soft state 
at the time of the radio observations that are plotted in Figure \ref{fig:f1}. The radio spectral index on 2011 May 30 was $\alpha = -0.02\pm0.12$, consistent with the presence of a compact jet at this epoch (see additional discussion in Section 4.2.1).

\begin{figure}[!t]
  \centering
 \includegraphics[width=0.5\textwidth]{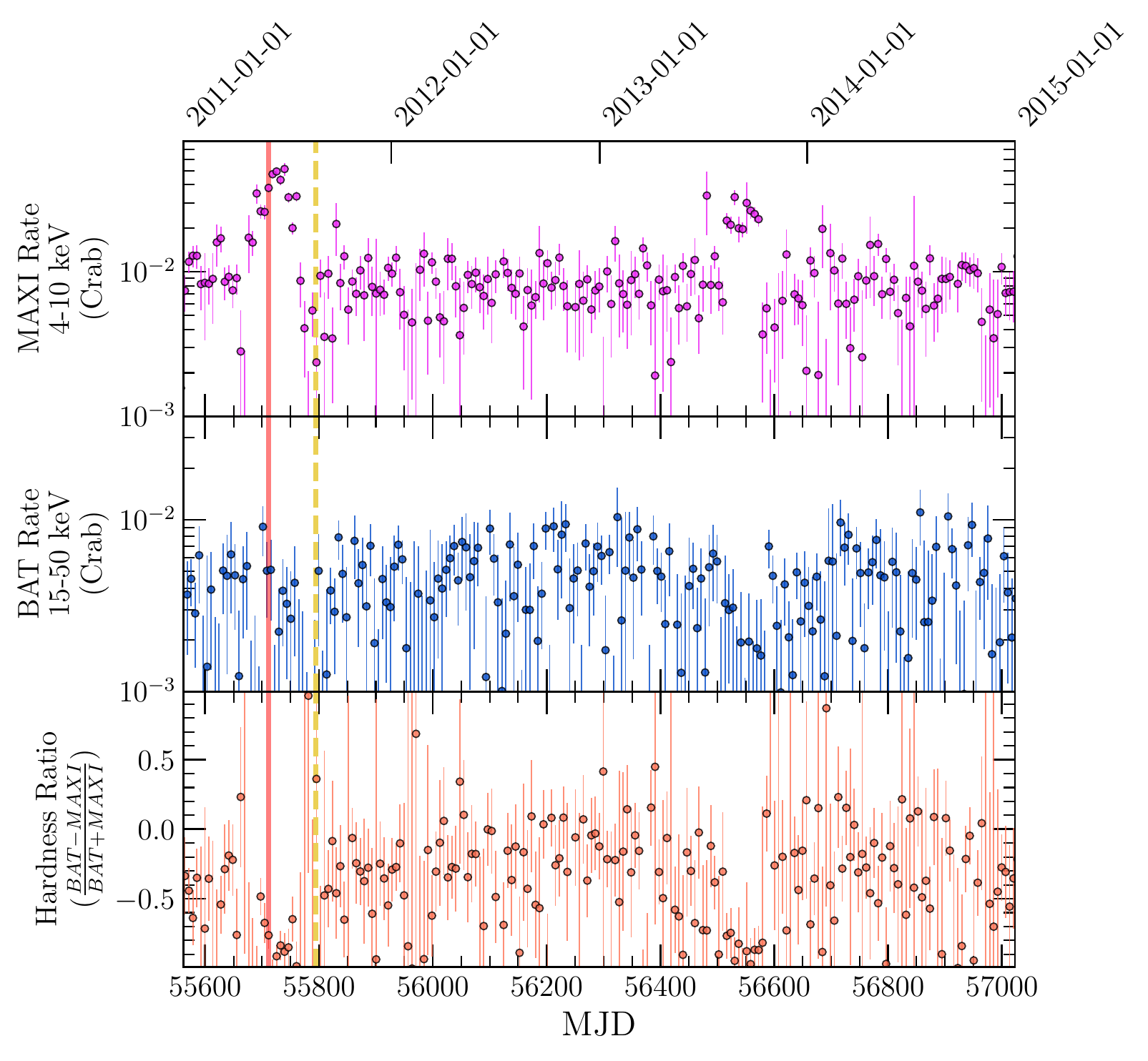}
  \caption{2011--2014 X-ray light curve of M15 (including flux from both AC 211 and M15 X-2) from MAXI (4--10 keV; top) and Swift/BAT (15--50 keV; middle), with hardness ratio in the bottom panel. The 2011 and 2013 flaring events are evidence. The epoch of our simultaneous 2011 May 30 VLA and Swift/XRT data is marked with a thick vertical salmon line. For reference, the second grouping of VLA data (2011 Aug 21/22) is marked with a dashed vertical yellow line; there are no simultaneous Chandra data for these epochs.}
  \label{fig:m15}
 \end{figure} 

For additional context, Figure \ref{fig:m15} shows the 2011--2014 MAXI and Swift/BAT X-ray light curves for M15. The 2011 and 2013 M15 X-2 flaring events are clearly visible in the MAXI X-ray light curve. The BAT data are noisy even when heavily binned, and AC 211 likely makes a substantial or even dominant contribution to the hard X-ray flux at all epochs, making it difficult to quantitatively interpret the plotted hardness ratio. Given the results of the Swift/XRT spectral analysis above, and that the 2011 May 30 radio data were obtained on the rise to peak, a reasonable interpretation is that these data were taken during a transition from a hard state to a softer one, with the jet not yet quenched.

\subsubsection{M15 X-3}

M15 X-3 is a so-called ``very faint X-ray transient", peaking at $\sim 10^{34}$ erg s$^{-1}$ \citep{Arnason15} rather than the value $\gtrsim 10^{36}$ erg s$^{-1}$ typical of LMXB transients. Even though it belongs to a different class than the other sources discussed in this paper, we include it for completeness given our use of deep radio continuum data for M15.

\citet{Arnason15} show that the secondary of M15 X-3 is likely a low-mass main sequence star but generally find no obvious explanation for the unusually faint X-ray outbursts, settling on propeller-mode accretion onto a rapidly rotating neutron star as perhaps the most likely explanation. They also consider the possibility that M15 X-3 could belong to the unusual class of ``transitional millisecond pulsars", which show flat-spectrum radio emission in their sub-luminous accretion disk states. The detection of flat-spectrum radio emission would be suggestive evidence that M15 X-3 could belong to this subclass of millisecond pulsars. 

We report a $3\sigma$ 5.0 GHz radio upper limit of $< 16.1$ $\mu$Jy on 2011 May 30, somewhat deeper than previously published radio continuum upper limits \citep{Heinke09b}. Quasi-simultaneous {Chandra} X-ray observations on 2011 May 31 yield an unabsorbed X-ray flux of $(3.5^{+1.4}_{-0.8}) \times 10^{-13}$ erg s$^{-1}$ cm$^{-2}$ \citep{Arnason15}, equivalent to $\sim 4\times 10^{33}$ erg s$^{-1}$ at the distance of M15.

Our quasi-simultaneous radio upper limit of $< 16.1$ $\mu$Jy is $\lesssim 10^{28}$ erg s$^{-1}$ at 5.0 GHz, which is not very constraining in the context of the known transitional millisecond pulsars or candidates, being comparable only to the detected radio emission in 3FGL J0427.9--6704 \citep{Li20}. If we assume the system stayed in a similar state for the longer period of 2011 May 22 to 2011 August 22, we can use the deeper, non-simultaneous radio data from \citet{Strader12b} to set a $3\sigma$ upper limit of $< 6.3$ $\mu$Jy ($\lesssim 4 \times 10^{27}$ erg s$^{-1}$ at 5.0 GHz). This limit is closer to the radio flux density observed for the transitional millisecond pulsar XSS J12270--4859 \citep{Hill11} but a factor of a few higher than that seen for the original transitional system PSR J1023+0038 \citep{Deller15}.

We conclude the existing radio data cannot strongly constrain whether M15 X-3 exhibits transitional millisecond pulsar-like radio properties, but that somewhat deeper data---perhaps challenging to obtain before the ngVLA era---would be interesting and useful.

\section{Discussion} 

\subsection{X-ray and Radio Luminosity: More Scatter Than Relation}

It is well-established that many black hole LMXBs accreting in the low/hard state (below a few percent of the Eddington rate; \citealt{Maccarone03}) show a consistent relationship between their luminosities in the X-ray ($L_X$) and the radio ($L_R$) close to $L_R \propto L_X^{0.6-0.7}$ (e.g., \citealt{Hannikainen98,Gallo03,Corbel13,Gallo18}, see also Figure \ref{fig:f1}). Not all black holes follow this relation. Some sit on steeper relations, which could be viewed as either being ``radiatively efficient" or with suppressed radio emission (e.g., \citealt{Coriat11,Ratti12,Huang14,Plotkin17,Carotenuto21,vandenEijnden21}).

\citet{Migliari06} collected the limited extant radio and X-ray measurements of neutron star X-ray binaries, suggesting that they typically produced jets, but (i) were fainter in the radio than black holes at the same $L_X$, and (ii) possibly followed a steeper radio/X-ray correlation than did black holes. Subsequent studies emerged slowly, since transient neutron star LMXBs often change states quickly and are intrinsically faint.

However, newly available facilities in the last decade have allowed for more successful X-ray and radio follow-up of neutron star LMXBs, especially in the more poorly studied regime $L_X < 10^{36}$ erg s$^{-1}$. To the extent that there is a unifying conclusion from these studies, it is that neutron star LMXBs show an enormous variety of behaviors
\citep{Tudor17,vandenEijnden21}. While it is possible to fit a mean radio/X-ray correlation to the hard state accreting neutron stars, which appears to have a slope similar to that for black holes but with a lower normalization (\citealt{Gallo18}, see also Figure \ref{fig:f1}), individual systems deviate from this mean relation to a much greater degree than for black holes. Some systems, such as the accreting millisecond pulsar IGR J16597--3704 and the ultracompact system 4U 1543--624, are much fainter in the radio than predicted from this mean relation \citep{Tetarenko18,Ludlam19}. Others, such as the accreting millisecond pulsar IGR J17591--2342, are likely as radio-loud as accreting black holes over some range in X-ray luminosity, perhaps reaching down to $L_X \sim 4\times 10^{35}$--$10^{36}$ erg s$^{-1}$ \citep{Russell18,Gusinskaia20b}. The deviations occur not only in normalization but in slope: there is good evidence that in at least in some systems, the radio emission becomes much fainter, or is quenched entirely, at $L_X \lesssim 10^{36}$ erg s$^{-1}$ 
\citep{Gusinskaia17,Gusinskaia20a}.

The results from this current paper, combining persistent systems with the outbursting binary GRS 1747--312, provide strong support for this picture of a broad range of behaviors for different neutron star LMXBs. Since these systems are all located in Galactic globular clusters, they also add the crucial element of known distances, which allows firm luminosity measurements that are lacking for many of the field systems.

Perhaps the only hard/island state accreting neutron star in our sample that that sits close to the mean radio/X-ray relation is the ultracompact 4U 1820--30, which does so at $L_X \sim 10^{37}$ erg s$^{-1}$ in Figure \ref{fig:f1}. M15 X-2 (also an ultracompact) appears to have similar radio/X-ray properties to 4U 1820--30 at this epoch, but the situation is slightly more complex, as the M15 X-2 observations may have occurred while the binary was transitioning to a softer state (Section 3.6.2).

GRS 1747--312 is a spectacular outlier. In the 2015 quasi-simultaneous radio/X-ray epoch, it is the most radio-luminous neutron star LMXB ever observed at $L_X < 10^{35}$ erg s$^{-1}$, sitting near the upper envelope of the black hole radio/X-ray correlation (Figure \ref{fig:f1}). In a subsequent quasi-simultaneous epoch in 2018, it has the lowest $L_X$ detection of radio emission for any ``normal" neutron star LMXB (i.e., excluding the transitional millisecond pulsars), again sitting close to the black hole radio/X-ray correlation. It shows substantial radio and X-ray variability between different epochs. In a third quasi-simultaneous epoch, it is not detected in the radio, suggesting at least a factor of 15 variability in the radio at these $L_X$ of a few $\times 10^{34}$ erg s$^{-1}$. As discussed below in Section 4.2.3, the radio variability and radio spectral index do not support the idea that the radio emission in GRS 1747--312 originates in a steady compact jet, and another physical mechanism is more likely.

XB 1832--330 is a persistent X-ray source, and its deep radio upper limit at $L_X \sim 2\times10^{35}$ is among the most constraining observed at this $L_X$, comparable to that seen in Aql X-1 during its fading after an outburst \citep{Gusinskaia20a}. As in Aql X-1, the XB 1832--330 data are consistent either with a steep radio/X-ray correlation for this binary or with a quenching of the jet at this luminosity. X1850--087 is another system in our sample with a deep constraining radio upper limit in the hard state at a relatively high $L_X \sim 10^{36}$ erg s$^{-1}$, but these data may have been taken near the start of an outburst, making their interpretation more challenging; we defer further discussion of this system to Section 4.2.3.

As an accretion disk corona source, the radio properties of AC 211 in M15 are also not straightforward to interpret, but if it is an accreting neutron star, it is rather radio-loud compared to other systems at this $L_X$. One potential explanation, discussed in Section 3.6.1, is that its luminosity is actually near the Eddington limit and its radio/X-ray properties are akin to Z sources such as Sco X-1. Another, less likely possibility is that AC 211 instead contains a black hole, but this is both a priori unlikely and difficult to prove.

A number of recent papers have tried to determine whether there is a simple explanation for the wide variation in radio properties of neutron stars, but have found that the most likely potential explanations, such as the spin rate or magnetic field of the neutron star, cannot readily account for the observed differences \citep{Migliari11,Tetarenko18,Tudor17,vandenEijnden21}. These globular cluster neutron star LMXBs stress this point even further by widening the range of observed behaviors. All of the hard state radio/X-ray measurements here are for ``normal" persistent or transient neutron stars; none of these systems are known to be accreting millisecond X-ray pulsars, which are suspected (but not proven) to have higher magnetic fields than other neutron star LMXBs \citep{Patruno21}. Hence these data are consistent with the idea that magnetic field strength is not the primary determinant of radio-loudness for accreting neutron stars.

\begin{figure*}
     \includegraphics[width=1.0\textwidth]{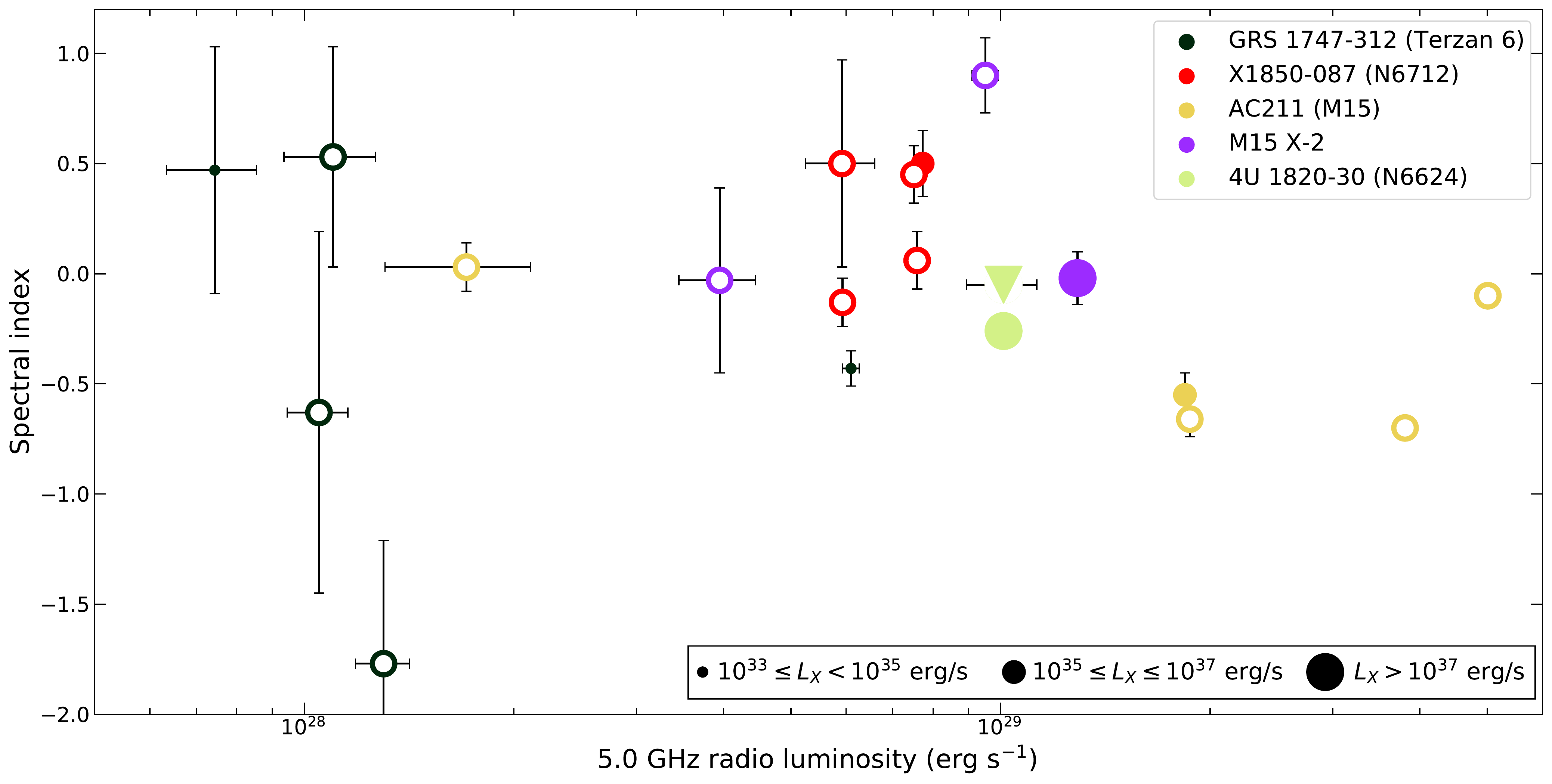}
    \caption{The relationship between radio spectral index and radio luminosity at 5.0 GHz.
 Filled markers have an associated quasi-simultaneous X-ray observation, while unfilled markers do not. The size of the marker indicates an estimate of the X-ray luminosity: small points $10^{33} \leq L_X \leq 10^{35}$ erg s$^{-1}$ , medium points $10^{35} \leq L_X \leq 10^{37}$ erg s$^{-1}$, and the largest points $L_X > 10^{37}$ erg s$^{-1}$. The upside down triangle represents a spectral index upper limit for one epoch of 4U 1820--30. Steady compact jets are expected to consistently show flat to mildly inverted spectral indices ($0 \lesssim \alpha \lesssim +0.5$). Strongly inverted spectral indices ($\alpha \gtrsim +0.5$) reflect more optically thick synchrotron emission, while more negative spectral indices (typically $-0.8 \lesssim \alpha \lesssim -0.4$) represent partially or entirely optically thin synchrotron. As discussed in the text, there may be a weak trend for more negative spectral index at the highest radio luminosity, but this is primarily driven by AC 211.}
    \label{fig:f5}
\end{figure*}

 \subsection{Spectral Indices}
The radio emission of neutron star LMXBs is thought to be non-thermal synchrotron radiation. Over some range of frequencies, synchrotron radiation is expected to emit as a power law, with $S_{\nu} \propto \nu^\alpha$.

The expected value of the spectral index depends on the origin of the synchrotron emission. Within the jet paradigm, flat ($\alpha \sim 0$) to inverted ($\alpha \gtrsim 0$) spectra are generally associated with an optically thick, steady, compact jet in the low/hard X-ray state \citep{Blandford79, Fender01}. Steeper ($\alpha \lesssim 0$) spectra suggest less optically thick emission, with $\alpha \sim -0.4$ to --0.8 expected for optically thin synchrotron, as might arise from the ejection of discrete jet blobs during hard to soft X-ray state transitions often observed in black hole LMXBs (e.g., \citealt{MillerJones12}). 

Only a handful of neutron star LMXBs have published high-quality estimates of the radio spectral index in the low/hard state, typically at $L_X \gtrsim 10^{36}$ erg s$^{-1}$. These are mostly consistent with the flat to inverted spectral index expected for an optically-thick compact jet \citep{Migliari10,MillerJones10,TetarenkoA16,Gusinskaia17,vandenEijnden18}. Possible exceptions are SAX J1808.4--3658, which had a mildly negative mean spectral index ($\alpha = -0.24\pm0.10$) at $L_X \sim 10^{36}$ erg s$^{-1}$ during its 2015 outburst \citep{Tudor17}, and IGR J17591--2342, which showed marginal evidence for evolution from a flat/inverted to slightly negative spectral index over the course of an outburst, declining from $L_X \sim 2\times 10^{36}$ to $\sim 4\times 10^{35}$ erg s$^{-1}$ \citep{Gusinskaia20b}. For the four IGR J17591--2342 radio measurements during the fading part of the outburst, the mean spectral index was $\alpha = -0.35\pm0.11$, compared to a mean spectral index of $\alpha = +0.17\pm0.09$ for the three radio measurements during the initial radio-bright stage of the outburst.

In this paper we have presented spectral index measurements for five globular cluster binaries, most with multiple measurements, representing a substantial increase in neutron star LMXBs with well-measured radio spectral indices. 

\subsubsection{4U1820--30 and M15 X-2}

For the 2015 ATCA observation of 4U 1820--30 we find $\alpha = -0.26\pm0.05$ in the hard state at $L_X \sim 10^{37}$ erg s$^{-1}$. This well-measured spectral index has a slightly steep value, which could suggest it is partially optically thin at these frequencies. In addition, the 5.0 GHz radio flux density is essentially identical to that observed with ATCA in 2014 when 4U1820--30 was in a brighter ($L_X \sim 8 \times 10^{37}$ erg s$^{-1}$) soft state \citep{diaztrigo2017}. We note that when the present paper was close to submission, a new paper appeared with a comprehensive analysis of new and archival radio continuum and X-ray observations of 4U 1820--30 \citep{Russell21}. Their results are consistent with, but more extensive than, the spectral index analysis in our paper, and suggest a transition from flat-spectrum steady compact jet emission in the low (island) state to steeper, possibly transient emission from jet ejecta in the high (banana) state.

The radio spectral index measurements for M15 X-2 all come from its X-ray brightening event in May 2011. While we do not have X-ray spectral information for the first radio epoch, the radio continuum measurements are consistent with the likely initial presence of a discrete optically thick synchrotron blob ($\alpha = +0.90\pm0.17$) which then fades before recovering in radio luminosity to a flat spectrum ($\alpha = -0.02\pm0.12$) consistent with a compact jet at $L_X \sim 10^{37}$ erg s$^{-1}$. As discussed in Section 3.6.2, at this latter epoch the system is consistent either being in a hard state or in a transition from a hard state to a soft state on the rise to the peak of the brightening event. We cannot definitively decide between these possibilities with these data: there is no evidence for the jet being quenched in the May 30 VLA observation. Overall, in May 2011 M15 X-2 shows ``classic" radio behavior for LMXB in the initial stages of an X-ray flare/outburst.

\subsubsection{AC 211: A Candidate Z Source}

The spectral index data for AC 211 are from the same dataset as M15 X-2. However, the AC 211 data not well-explained by a standard
hard/island state accreting neutron star model. In the first two observations (2011 May 22 and 26) the source shows a flat spectral index, consistent with a jet, while its radio luminosity increased by a factor of $\sim 3$ between the first and second epochs. On 2011 May 30 (four days later), a steeper spectrum is observed ($\alpha = -0.55\pm0.10$) despite little change in the 5.0 GHz flux density. The next radio data available are three months later, on 2011 August 21, with both the spectrum and flux matching the previous epoch. But only one day later (2011 August 22) the 5.0 GHz flux doubles while the spectrum remains steep. To summarize, AC 211 shows substantial variation in both its radio luminosity and spectral index, but with no clear correlation between these. 

Previously published observations of AC 211 also show evidence for variability. A pre-upgrade VLA image of M15 found a 1.4 GHz source with a flux density of $\sim 1.8$ mJy at a position consistent with AC 211 \citep{Kulkarni90}; this measurement is roughly consistent with the brightest 5.0 GHz flux density we see, assuming an $\alpha=-0.7$ spectral index. However, very long baseline imaging finds a typical flux density of around 200 $\mu$Jy at 1.6 GHz \citep{Kirsten14}, which would be more consistent with a flat spectral index and a fainter overall flux level. An interpretation of AC 211 as a Z source accreting at close to the Eddington limit would give a straightforward explanation for the variability in the radio flux density and spectral index as due to discrete cases of jet ejection events (e.g., \citealt{Hjellming90, Fomalont01}). This interpretation appears to offer a more plausible explanation for the high radio luminosity of AC 211 than the speculative idea that the primary is a black hole.

\subsubsection{The Weirdos: X1850--087 and GRS 1747--312}

For X1850--087 the interpretation of its spectral indices are mixed together with our interpretation of its fast radio variability, which is somewhat confusing (Section 3.2). The strongly inverted to flat spectral indices observed from 2014 May 5--9 appear to coincide with an X-ray flare, so can be explained as transient optically thick synchrotron emission. However, the source is undetected in the radio on 2014 May 13 before becoming bright again, with a flat to inverted spectral index, on May 18--20. Since the system was not detected in the radio in an apparently normal hard state ($L_X \sim 10^{36}$ erg s$^{-1}$) on 2014 April 5, one possibility is that this system has no jet in the hard state at this $L_X$, and that it transitioned to the hard (or intermediate hard) state by May 13, before undergoing a fast reflare by May 18 associated with partially optically thick synchrotron emission. This scenario is not generally consistent with previous observations of X1850--087 in the X-ray and radio (Section 3.2) but cannot be ruled out either. The level of X-ray variability observed in X1850--087 and other persistent ultracompact LMXBs is already challenging to explain (e.g., \citealt{intZand07,Maccarone10,Cartwright13}), and these new radio results are intriguing, highlighting the need for more coordinated radio and X-ray observations of this source.

The final source with spectral index measurements is GRS 1747--312. It was well-detected in many radio epochs, despite being relatively faint with $L_X \sim 7\times10^{33}$ to $4\times10^{34}$ erg s$^{-1}$. At the first (and radio-brightest) epoch in 2015, sitting at the upper edge of the black hole radio/X-ray correlation, it had $\alpha = -0.43\pm0.08$ in ATCA data, more consistent with optically thin synchrotron than the flat/inverted emission associated with a steady jet. In an average of the 2018 VLA detections, it was fainter, with a mean $\alpha = -0.51\pm0.29$, consistent with the previous ATCA epoch. However, the range of values measured in the individual epochs was enormous ($-1.77\pm0.56$ to $+0.53\pm0.50$) and consistent with an intrinsic spread in the spectral indices, though this cannot be proven with these data due to the large uncertainties on the individual measurements.

The radio spectral indices and extreme radio variability are not consistent with expectations for a steady compact jet. One possibility is that the radio emission is associated with individual luminous discrete ejecta events; in principle it might be possible to image the brightest events with very long baseline interferometry. Other explanations may also be feasible. For example, the transitional millisecond pulsar PSR J1023+0038 is radio-loud in quiescence with a variable spectral index that changes on short timescales, though it is not as radio-loud as GRS 1747--312, and typically has a flatter/inverted spectral index \citep{Deller15}. For PSR J1023+0038, the radio emission likely comes from non-steady synchrotron bubbles created near the interface between the inner disk and the neutron star \citep{Bogdanov18}. PSR J1023+0038 is not alone: other confirmed and candidate transitional systems show luminous radio emission in their sub-luminous disk states \citep{Hill11,Jaodand19,Li20}. It is not clear whether the same physical mechanism powers the radio emission in all of these systems, and a more ``standard" propeller mechanism, producing a radio outflow, could instead be at play in a subset of them---or in GRS 1747--312.

\subsubsection{Spectral Index and Radio Luminosity}

In Figure \ref{fig:f5}, we show the spectral index of our sources as a function of their 5.0 GHz radio luminosity. More sources are plotted here than in Figure \ref{fig:f1}, since we do not require a simultaneous X-ray measurement to plot them here. Epochs with only radio upper limits are not plotted, since these have no spectral index measurements or constraints.

As discussed for the individual sources above, there is no clear, strong relationship between radio luminosity and spectral index for individual sources or for the sample as a whole. There may be weak evidence for a change at the highest radio luminosities: those with $L_R > 10^{29}$ erg s$^{-1}$ have a mean $\alpha = -0.39 \pm 0.02$, while those below this radio luminosity have a mean $\alpha = +0.01 \pm 0.04$. However, this result is not very robust as it is dominated by the unusual source AC 211 at high luminosities. We also do not see evidence that the relationship between radio luminosity and spectral index varies substantially as a function of broad bins of X-ray luminosity. However, this is also not straightforward to interpret, since we do not have simultaneous X-ray data (and hence proper spectral classification) for each measurement.

\section{Summary} 
In this paper, we used X-ray and radio data to investigate the relationship between the accretion flow and jet/outflow in six persistently accreting neutron star LMXBs and two other transient sources in Galactic globular clusters. These data represent the largest sample of quasi-simultaneous radio and X-ray observations of persistently accreting neutron star LMXBs in the low/hard state. The location of these sources in the standard radio/X-ray diagram for accreting compact objects broadly follows the results from previous studies of neutron star LMXBs, but with greater extremes of luminous radio emission and constraining upper limits. From the current sample alone there is no evidence for a well-defined correlation between X-ray and radio emission for neutron star LMXBs, but instead a large scatter in properties at all observed $L_X$ values. Nearly all of the sources in our sample with multiple measurements of their radio properties show unusual variability in both their luminosity and radio spectral index. This highlights the need for additional, high-cadence simultaneous radio and X-ray observations of neutron stars, even in nominally persistent systems, to make progress in understanding jets/outflows from accreting neutron stars.

\section{Acknowledgements} 
We would like to acknowledge the thoughtful comments of an anonymous referee that substantially improved the paper.

TP dedicates this paper to the essential workers and cleaning staff at Michigan State and other institutions around the world, whose labor provided us with crucial resources and a space where astrophysical research can thrive during the time of COVID-19.

We acknowledge support from NSF grant AST-1308124 and NASA grants NNX16AN73G,  Chandra-GO7-18032A, Chandra-GO8-19122X, and 80NSSC21K0628. TP is funded by the NSF Graduate Research Fellowship (DGE 1848739). JS acknowledges support from the Packard Foundation. COH is supported by NSERC Discovery Grant RGPIN-2016-04602. GRS is supported by NSERC Discovery Grants RGPIN-2016-06569 and RGPIN-2021-04001.

The National Radio Astronomy Observatory is a facility of the National Science Foundation operated under cooperative agreement by Associated Universities, Inc. The Australia Telescope Compact Array is part of the Australia Telescope National Facility which is funded by the Australian Government for operation as a National Facility managed by CSIRO. The scientific results reported in this article are partially based on on observations made by the Chandra X-ray Observatory. This work made use of data supplied by the UK Swift Science Data Centre at the University of Leicester. This research has made use of data and software provided by the High Energy Astrophysics Science Archive Research Center (HEASARC), which is a service of the Astrophysics Science Division at NASA/GSFC and the High Energy Astrophysics Division of the Smithsonian Astrophysical Observatory.\\

\noindent \textit{Facilities}: VLA, ATCA, Swift/XRT, Chandra

\noindent \textit{Software}: Astropy \citep{Robitaille13}, AIPS \citep{Greisen03}, \textsc{CASA} \citep{McMullin07}, \textsc{CIAO} \citep{Fruscione06}, \textsc{Heasoft} \citep{HEASARC14} Matplotlib \citep{Hunter07}, Miriad \citep{Sault95}, NumPy \citep{harris20}, pandas \citep{Mckinney10}, \textsc{Xspec} \citep{Arnaud96}

\bibliography{references.bib}

\newpage
\begin{deluxetable*}{lllllrlr}
\label{tab:lrlx}
\tabletypesize{\tiny}
\tablecaption{Quasi-Simultaneous Radio and X-ray Observations} 
\label{tab:quasisim}
\tablehead{
\colhead{ID} & 
\colhead{Cluster}  & 
\colhead{R.A.\tablenotemark{a}}  & 
\colhead{Dec.\tablenotemark{a}}     &
\colhead{Date of Radio Obs. (UT)}&
\colhead{$S_{\textrm{5.0 GHz}}$\tablenotemark{b}}&
\colhead{Date of X-ray Obs. (UT)}&
\colhead{X-ray Flux\tablenotemark{c}}                \\
                & 
     &
(h:m:s)             &                   
($^{\circ}$:\arcmin:\arcsec)  &
             &
($\mu$Jy)      &
             &
(ergs s$^{-1}$cm$^{-2}$) }
\startdata
4U 1746--37 & NGC\,6441 & 17:50:12.695 & --37:03:06.56 & 2015 April 15.250 (14.542--15.958) & $< 13.7$\tablenotemark{e} & 2015 April 13.045 (13.027--13.063) & $(8.6\pm2.0) \times10^{-10}$ \\
GRS 1747--312 & Terzan 6 & 17:50:46.912\tablenotemark{d} & --31:16:29.18\tablenotemark{d} & 2015 April 18.250 (17.542--18.958)  & $221\pm6.4$ & 2015 April 16.690 (16.672--16.709) & $(5.3^{+4.0}_{-2.4}) \times 10^{-12}$ \\
              &          &                               &               & 2018 March 25.594 (25.557--25.630) & $< 12.9$ & 2018 March 26.567 (26.549--26.586) & $(8.0^{+13.6}_{-4.8}) \times 10^{-12}$ \\
                &               &               &                   & 2018 April 30.380 (30.344--30.417) & $26.9\pm4.3$ & 2018 May 01.569 (01.550--01.587) &  $(1.3^{+1.2}_{-0.8}) \times 10^{-12}$ \\
4U 1820--30 & NGC\,6624 & 18:23:40.499 & --30:21:40.10 & 2014 July 17.521 (17.417--17.625) & $241\pm28$ & 2014 July 19.492 (19.459--19.524) & $(9.6\pm0.2) \times10^{-9}$ \\
            &          &               &                & 2015 April 25.250 (24.542--25.958) & $241\pm5.1$ & 2015 April 24.861 (24.810--24.912) & $(1.5\pm0.1) \times10^{-9}$ \\
XB 1832--330 & NGC\,6652 & 18:35:43.656 & --32:59:26.35 & 2017 May 22.415 (22.350--480) & $<6.6$\tablenotemark{e} & 2017 May 22.783 (22.343--23.223) & $(1.9\pm0.1) \times10^{-11}$ \\
X1850--087 & NGC\,6712 & 18:53:04.867 & --08:42:20.34 & 2014 April 05.626 (05.590--05.661) & $<$12.9 & 2014 April 06.463 (06.424-06.502) & $(1.6\pm0.1) \times10^{-10}$ \\   
AC 211 & M15 & 21:29:58.310 & 12:10:02.66 & 2011 May 30.512 (30.455--30.568) & $292\pm6.3$          & 2011 May 31.017 (30.691--31.343) & $(4.46\pm0.05) \times10^{-10}$ \\
M15 X-2 & M15 & 21:29:58.132 & 12:10:02.24 & 2011 May 30.512 (30.455--30.568) & $205\pm6.3$ & 2011 May 31.017 (30.691--31.343) &  $(9.46\pm0.03) \times10^{-10}$  \\
M15 X-3 & M15 & 21:29:58.161 & 12:09:39.93 & 2011 May 30.512 (30.455--30.568) & $<$16.1 & 2011 May 31.017 (30.691--31.343) &  $3.5^{+1.4}_{-0.8}~\times10^{-13}$ \\
\enddata
\tablenotetext{a}{ICRS coordinates.}
\tablenotetext{b}{Radio flux density at 5.0 GHz.}
\tablenotetext{c}{Unabsorbed X-ray flux from 1--10 keV.}
\tablenotetext{d}{This is an average of the VLA 5.0 and 7.1 GHz positions.}
\tablenotetext{e}{Assuming $\alpha = 0$.}
\end{deluxetable*}

\begin{deluxetable*}{lrrrrr} 
\label{tab:x1850}
\tabletypesize{\footnotesize}
\tablewidth{0pt} 
\tablecaption{VLA Radio Observations of X1850--087} 
\tablehead{Date & $\nu_{\textrm{low}}$ & $S_{\textrm{low}}$ &  $\nu_{\textrm{high}}$ & $S_{\textrm{high}}$ & $\alpha$ \\
  (UT) & (GHz) &  ($\mu$Jy) &(GHz) & ($\mu$Jy) &  }
\startdata
1989 February 19.589 (14.478--14.700) & 4.9 & $156\pm24$\tablenotemark{a} & \nodata & \nodata & \nodata \\
1991 December 15.834 (15.683--15.985) & 4.9 & $134\pm23$\tablenotemark{a} & \nodata & \nodata & \nodata \\
1998 September 29.083 (29.061--29.105) & 4.9 & $< 135$\tablenotemark{a} & \nodata & \nodata  & \nodata \\
2014 April 05.626 (05.590--05.661) & 5.0 & $< 12.9$ & 7.4 &  $< 12.3$ & \nodata \\
2014 May 05.600 (05.584--05.615) & 5.0 & $167.0\pm8.0$  & 7.4 &  $202.9\pm6.6$ & $+0.50\pm0.15$ \\
2014 May 08.588 (08.552--08.623) & 5.0 & $164.0\pm5.0$  & 7.4 &  $155.7\pm4.6$ & --$0.13\pm0.11$ \\
2014 May 09.497 (09.482--09.512) & 5.0 & $165.3\pm6.8$  & 7.4 & $197.0\pm6.2$ & $+0.45\pm0.13$ \\
2014 May 13.574 (13.539--13.610) & 5.0 & $<14.0$ & 7.4 & $<12.9$ & \nodata \\
2014 May 18.492 (18.476--18.507)	& 5.0 & $194.2\pm7.0$ & 7.4 & $199.1\pm7.2$ & $+0.06\pm0.13$ \\
2014 May 20.581 (20.566--20.596)	& 5.0 & $127.7\pm18.7$ & 7.4 & $155.3\pm17.7$ & $+0.50\pm0.47$ \\
 \enddata
\tablenotetext{a}{Central frequency 4.9 GHz, not 5.0 GHz.}
\end{deluxetable*}

\begin{deluxetable*}{lrrrrr} 
\tabletypesize{\small}
\tablewidth{0pt} 
\tablecaption{Radio Observations of GRS 1747--312} \tablehead{Date & $\nu_{\textrm{low}}$ & $S_{\textrm{low}}$ & $\nu_{\textrm{high}}$ & $S_{\textrm{high}}$ & $\alpha$ \\
  (UT) & (GHz) & ($\mu$Jy) & (GHz) & ($\mu$Jy) &  }
\label{table:grstable}
\startdata
2015 April 18.250 (17.542--18.958) & ATCA/5.5 & $213.3\pm5.1$\tablenotemark{a} & ATCA/9.0 & $172.4\pm5.2$\tablenotemark{b} & --$0.43\pm0.08$ \\
2018 March 25.594 (25.557--25.630) & VLA/5.0 & $< 12.9$ & VLA/7.1 & $< 11.7$ & \nodata \\
2018 April 30.380 (30.344--30.417) & VLA/5.0 & $26.9\pm4.3$ & VLA/7.1 & $32.1\pm4.1$ & $+0.47\pm0.56$ \\
2018 May 21.307 (21.271--21.344) & VLA/5.0 & $47.4\pm4.2$ & VLA/7.1 & $26.0\pm4.3$ & --$1.77\pm0.56$ \\
2018 May 31.416 (31.401--31.432) & VLA/5.0 & $42.5\pm6.0$ & VLA/7.1 & $51.3\pm6.0$ & $+0.53\pm0.50$ \\
2018 June 03.270 (03.255--03.285) & VLA/5.0 & $38.4\pm6.4$ & VLA/7.1 & $31.5\pm6.7$ & --$0.63\pm0.82$ \\
2018 April 30--June 3\tablenotemark{c} & VLA/5.0 & $38.4\pm2.3$\tablenotemark{d} & VLA/7.1 & $32.2\pm2.5$\tablenotemark{e} & --$0.51\pm0.29$ \\
\enddata
\tablenotetext{a}{ATCA/5.5 GHz ICRS position of 17:50:46.9139(122) --31:16:29.04(42).}
\tablenotetext{b}{ATCA/9.0 GHz ICRS position of 17:50:46.9145(82) --31:16:29.14(28).}
\tablenotetext{c}{A stack of the four detected VLA epochs in the $uv$ plane.}
\tablenotetext{d}{VLA/5.0 GHz ICRS position in stack of 17:50:46.9119(35) --31:16:29.202(91).}
\tablenotetext{e}{VLA/7.1 GHz ICRS position in stack of 17:50:46.9112(25) --31:16:29.159(66).}
\end{deluxetable*}

\begin{deluxetable*}{llrrr}
\tablecaption{VLA Observations of AC 211 and M15 X-2}
\label{tab:m15}
\tablehead{Source & Date & $S_{5.0 \textrm{GHz}}$ &  $S_{7.4 \textrm{GHz}}$  & $\alpha$ \\
                  & (UT) & ($\mu$Jy) &  ($\mu$Jy) &   }
 \startdata
 AC 211 & 2011 May 22.535 (22.521--22.549) & $270.8\pm6.4$ &  $273.8\pm7.5$\tablenotemark{a} &  $+0.03\pm0.11$  \\
      & 2011 May 26.503 (26.488--26.518) & $794.2\pm7.9$ & $764.8\pm6.4$ &  $-0.10\pm0.03$  \\
      & 2011 May 30.512 (30.455--30.568) & $292.0\pm6.3$  & $235.5\pm7.3$ & $-0.55\pm0.10$ \\
      & 2011 August 21.377 (21.291--21.383) & $295.9\pm5.7$  &  $228.7\pm6.0$ &  $-0.66\pm0.08$ \\
      & 2011 August 22.438 (22.392--22.485) & $602.5\pm5.2$  & $458.3\pm4.8$ &  $-0.70\pm0.03$  \\
 M15 X-2 & 2011 May 22.535 (22.521--22.549) & $150.5\pm6.4$  & $203.9\pm7.5$\tablenotemark{a} &  $+0.90\pm0.17$ \\
         & 2011 May 26.503 (26.488--26.518) & $63.6\pm7.9$  & $62.6\pm6.4$ & $-0.03\pm0.42$ \\
        & 2011 May 30.512 (30.455--30.568) & $205.0\pm6.3$  & $203.7\pm7.3$ & $-0.02\pm0.12$ \\
         & 2011 August 21.377 (21.291--21.383) & $<$18.3 &  $<$18.0 &  \nodata \\
         & 2011 August 22.438 (22.392--22.485) & $<$17.6 & $<$15.1 & \nodata \\
 \enddata
\tablenotetext{a}{Central frequency of 7.0 GHz, not 7.4 GHz.}
  \end{deluxetable*}

\appendix
\section{Radio Continuum Images}

\restartappendixnumbering

\begin{figure*}
    \includegraphics[width=1.0\textwidth]{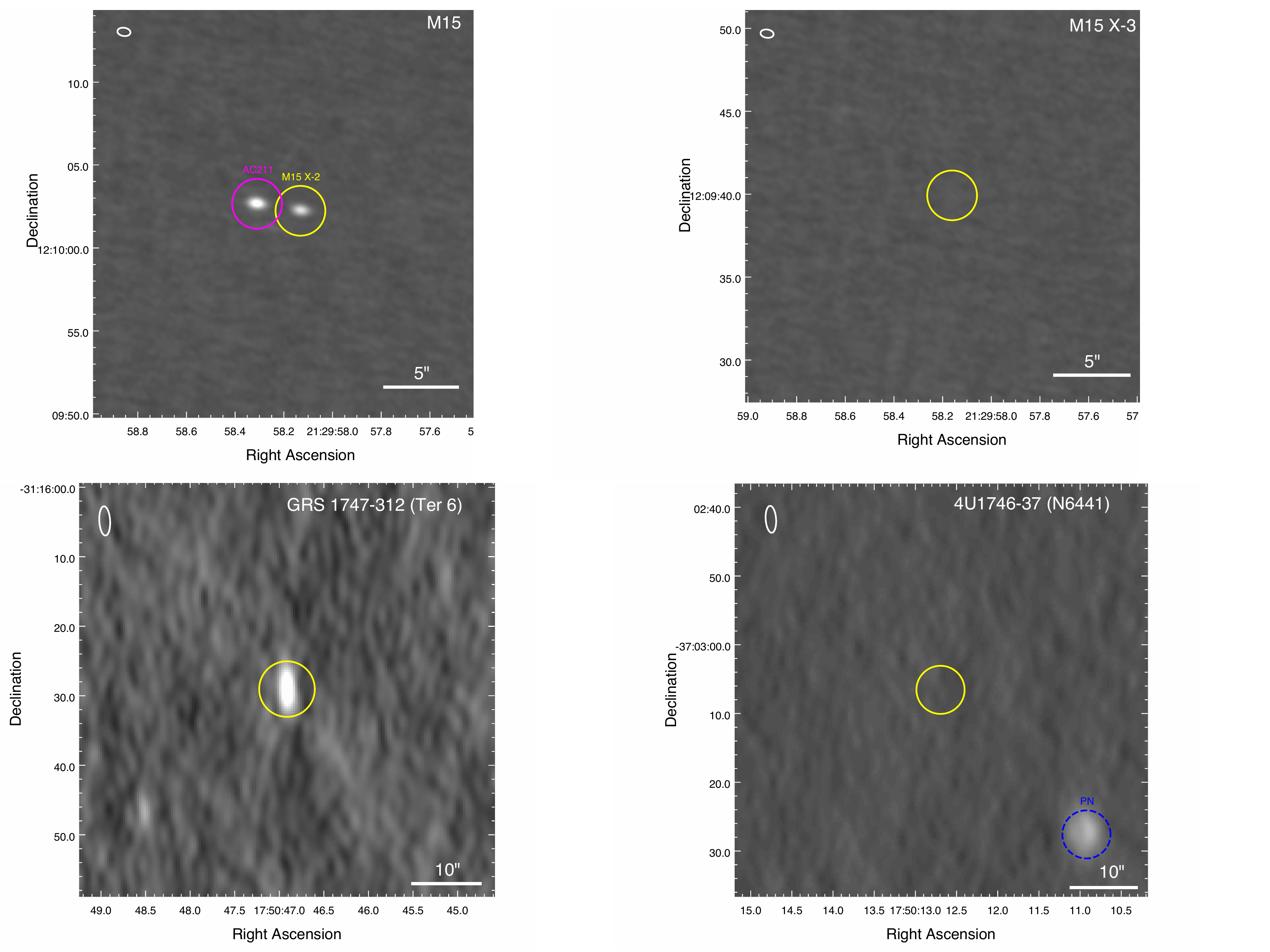}
    \caption{Representative radio continuum images of AC 211, M15 X-2, M15 X-3, GRS 1747--312, and 4U 1746--37. Top Left Panel: VLA 5.0 GHz image of M15 X-2 and AC 211, marked in yellow and magenta circles respectively. Top Right Panel: VLA 5.0 GHz image of the location of M15 X-3 (the source is undetected, with a $3\sigma$ upper limit of $< 16.1$ $\mu$Jy). Bottom Left Panel: ATCA 5.5 GHz image of GRS 1747--312 from 2015 April 18. Bottom Right Panel: ATCA 5.5 GHz image of the location of 4U 1746--37 (yellow); the source is undetected with a $3\sigma$ upper limit of $< 13.7$ $\mu$Jy. Here the visible source (circled in blue) is the planetary nebula JaFu 2 \citep{Bond20}. In the top left corner of each panel, the synthesized beam is shown in white.}
    \label{fig:fim1}
\end{figure*}

\newpage
\begin{figure*}
    \includegraphics[width=1.0\textwidth]{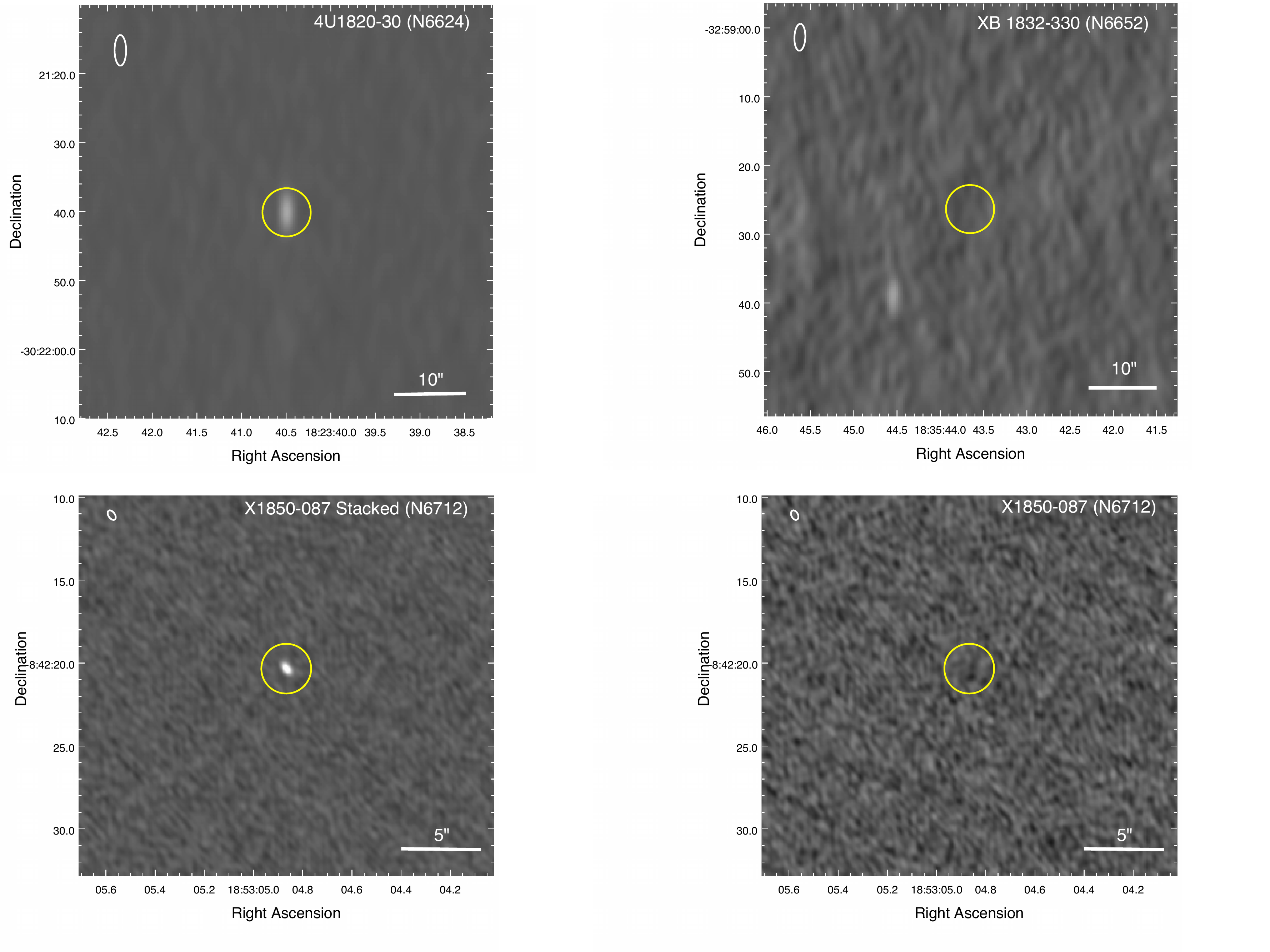}
    \caption{Representative radio continuum images of 4U1832--30, XB1832--330, and X1850--087. Top Left Panel: ATCA 5.5 GHz image of 4U1832--30. Top Right Panel: VLA 10 GHz image of XB1832--330. The source is undetected with a $3\sigma$ upper limit of $< 6.6$ $\mu$Jy (the faint source to the southwest is the transitional millisecond pulsar candidate NGC 6652B; \citealt{Paduano21}). Bottom Left Panel: VLA 5.0 GHz May 2014 stacked image of X1850--087, with the source well-detected. Bottom Right Panel: VLA 5.0 GHz image of X1850-087 from 2014 April 5. The source is undetected with a $3\sigma$ upper limit of $< 12.9$ $\mu$Jy.  Synthesized beams are shown in white at the top left corner in every image.}
    \label{fig:fim2}
\end{figure*}

\end{document}